\documentclass[]{aastex631}

	\usepackage{graphicx}	% Including figure files
	\usepackage{amsmath}
	\usepackage{amssymb}	% Extra maths symbols
	\usepackage{longtable}
	\usepackage{threeparttable}
	\usepackage{multirow}
	\usepackage{booktabs}
	\usepackage{array}
    \usepackage[figuresright]{rotating}
    \usepackage[T1]{fontenc}

	\shorttitle{Karachurin 12}
	\shortauthors{Sun et al.}
	\graphicspath{{./}{figures/}}

\begin{document}
			\title{A New IW And-Type Star: Karachurin 12 with Tilted Disks and Diverse cycles}

			%\correspondingauthor{August Muench}
			%\email{greg.schwarz@aas.org, gus.muench@aas.org}

\correspondingauthor{Sheng-Bang Qian}
\email{qiansb@ynu.edu.cn}	
\author{Qi-Bin Sun}
\affiliation{Department of Astronomy, School of Physics and Astronomy, Yunnan University, Kunming 650091, China}
\affiliation{Key Laboratory of Astroparticle Physics of Yunnan Province, Yunnan University, Kunming 650091, China}

\author{Sheng-Bang Qian}

\affiliation{Department of Astronomy, School of Physics and Astronomy, Yunnan University, Kunming 650091, China}
\affiliation{Key Laboratory of Astroparticle Physics of Yunnan Province, Yunnan University, Kunming 650091, China}

\author{Li-Ying Zhu}
\affiliation{Yunnan Observatories, Chinese Academy of Sciences, Kunming 650216, China}
\affiliation{University of Chinese Academy of Sciences, No.19(A) Yuquan Road, Shijingshan District, Beijing, China}

\author{Qin-Mei Li}
\affiliation{Department of Physics, college of physics, Guizhou university, Guiyang, 550025, China}

\author{Fu-Xing Li}
\affiliation{Department of Astronomy, School of Physics and Astronomy, Yunnan University, Kunming 650091, China}
\affiliation{Yunnan Observatories, Chinese Academy of Sciences, Kunming 650216, China}

\author{Min-Yu Li}
\affiliation{Yunnan Observatories, Chinese Academy of Sciences, Kunming 650216, China}
\affiliation{University of Chinese Academy of Sciences, No.19(A) Yuquan Road, Shijingshan District, Beijing, China}	

\author{Ping Li}
\affiliation{Yunnan Observatories, Chinese Academy of Sciences, Kunming 650216, China}
\affiliation{University of Chinese Academy of Sciences, No.19(A) Yuquan Road, Shijingshan District, Beijing, China}	

			%%%%%%%%%%%%%%%%%%%%%%%%%%%%%%%%%%%%abstract%%%%%%%%%%%%%%%%%%%%%%%%%%%%%%%%%%%%%%%

			\begin{abstract}

The IW And-type phenomenon in cataclysmic variables presents a significant challenge to the accretion disk instability model. Using photometric data from the All-Sky Automated Survey for Supernovae, the Zwicky Transient Facility, and the Transiting Exoplanet Survey Satellite, we identify Karachurin 12 as a new non-eclipsing IW And-type object with a cycle period of 35.69(3) days. We also report for the first time that Karachurin 12 is a negative superhump (NSH) system featuring a precessing tilted disks, with precession, orbital, and NSH periods of 4.9588(2) days, 0.3168895(13) days, and 0.2979861(8) days, respectively. Our analysis, using dips as index and NSHs as probe, reveals diverse cycle patterns in Karachurin 12, with NSH amplitude varying throughout the cycle. These findings offer new insights for studying tilted disks and the IW And-type phenomenon.	The mass-transfer burst model has difficulty explaining the observed variations in NSH amplitude, especially given the uncertainty surrounding the origin of the mass transfer burst. Meanwhile, the tilted thermally unstable disk model indicates a possible connection to the IW And-type phenomenon, but it also struggles to account for the detailed variations in Karachurin 12. Therefore, a wider range of factors must be considered to fully understand the complex changes in Karachurin 12.

\end{abstract}


%%%%%%%%%%%%%%%%%%%%%%%%%%%%%%%%%%%%Introduction%%%%%%%%%%%%%%%%%%%%%%%%%%%%%%%%%%%%%%%

\section{Introduction} \label{sec:intro}

Cataclysmic variables (CVs) are of considerable interest due to their unique characteristics and their contributions to astrophysical research. These systems typically consist of a white dwarf (the primary) and a late main-sequence star (the secondary). Key types of CVs include classical novae, recurrent novae, novae-like stars (NLs), dwarf novae (DNe), and magnetic CVs \citep{warner1995cat}. In these binary systems, the secondary star overflows its Roche lobe and transfers mass to the primary, forming a semi-detached close binary system.
When the magnetic field of the white dwarf is weak (below 1 MG), it allows for the formation of accretion disks around the white dwarf. In contrast, a stronger magnetic field (exceeding 1 MG) disrupts the accretion disks, leading to the formation of accretion curtains or columns (\citealp{1991MNRAS.248..233H,1995ASPC...85..185H,1999ApJ...519..324H,1996PASA...13...87F}).

Dwarf novae (DNe) represent a subclass of CVs, which are weakly magnetic or non-magnetic CVs. In these systems, the variability is primarily driven by the accretion disks, which cause outbursts that, while less intense than those seen in novae (which involve thermonuclear reactions), occur more frequently. DNe are typically categorized into three main types: Z Camelopardalis, SU Ursae Majoris, and U Geminorum. Normal DN outbursts result from luminosity variations driven by thermal instabilities in the accretion disks. This behavior is commonly explained by the accretion disk instability model (DIM; \citealp{1974PASJ...26..429O,1996PASP..108...39O,lasota2001disc,2017A&A...602A.102H}).

DIM posits that the accretion disk undergoes transitions between two stable states—cold and hot—driven by changes in opacity with temperature. In the low-temperature state where hydrogen is neutral (below $ \sim 6000 $ K), the accretion disk is stable and exhibits low viscosity. As the temperature increases and hydrogen becomes partially ionized, the disk becomes hotter, more viscous, and thermally unstable. When the temperature rises sufficiently to fully ionize hydrogen (above $ \sim 8000 $ K), the disk stabilizes again, at which point it has high viscosity, and its thermal behavior can be described by a classical S-shaped stability curve (See e.g., \citealp{lasota2001disc}; \citealp{dubus2018testing}; \citealp{2020AdSpR..66.1004H} for details).

 Z Camelopardalis (Z Cam) are particularly distinguished by their “standstill” behavior during the decline phase of outbursts, where their brightness stabilizes approximately 0.7 magnitudes below the peak level. DIM explains typical “standstill” by describing the disks as being in a hot stable state. When the mass transfer rate from the secondary star exceeds a critical value ($\dot{M}_{\text{crit}}$), the disks remain in a hot stable state, similar to that observed in Z Cam and NLs. Z Cam have mass transfer rates close to $\dot{M}_{\text{crit}}$, with slight variations potentially triggering the “standstill” phenomenon \citep{Meyer1983A&A, Smak1983ApJ, Honeycutt1998PASP}.
 However, during the “Z CamPaign” observation campaign, \cite{2011JAVSO..39...66S} observed for the first time that the “standstill” in IW And and V513 Cas did not conclude by returning to the quiescent state but instead culminated in an outburst, followed immediately by a dip and then a rapid return to “standstill”. This unusual behavior was identified by \cite{2013PASP..125.1421S}. \cite{Kato2019PASJ} categorized these objects as “IW And-type objects (or phenomena)” and proposed the existence of a previously unknown type of limit-cycle oscillation in IW And-type stars. Multiple IW And-type objects have been found in subsequent studies, such as V507 Cyg \citep{2019PASJThree}, ST Cha \citep{2021STCha}, IM Eri \citep{2020PASJIMEri}, KIC 9406652 \citep{KimuraKIC94066522020PASJ}and HO Pup \citep{2021ApJHOPuppis}.
 
IW And-type phenomenon is a new challenge to DIM, and the specific physical processes about it are still under debate. 
\cite{2014mass-transferoutbursts} was one of the first to systematically explanation the IW And-type phenomenon, suggesting that mass-transfer burst is the primary cause and partially reproducing the observed effects. However, the exact trigger for these mass-transfer oburst remains unresolved.
\citet{kimura2020thermal} conducted numerical simulations based on the tilted thermal-viscous instability model, which also successfully reproduces part of the IW And-type phenomenon. They suggest that a tilted accretion disks allow mass from the secondary star to more easily reach the inner disks, establishing a new cycle. In this cycle, the inner disks remain nearly always in a hot state, while the outer disk undergoes repeated outbursts, producing the observed light curve characteristics of the IW And-type phenomenon. As recognized by \citet{kimura2020thermal}, while some phenomena of IW And objects are reproduced, the modeled light curves differ in several aspects from the observed ones, including the timing of dips, the amplitude of brightening, the oscillation amplitudes, and the occurrence of Z Cam-type standstills (see \citealp{kimura2020thermal} for details)

The precession of tilted accretion disks has been observed across various types of celestial systems \citep[e.g.,][]{Foulkes2010MNRAS,Giacconi1973ApJ,2017MNRASAGN,2023NaturM87} and is particularly common in CVs, where the periods are typically only a few days \citep[e.g.,][]{2023MNRAS.520.3355S, sun2022study, Bruch2023}. In CVs, tilted accretion disks often exhibit signals known as “negative superhumps” (NSHs), with periods approximately 5\% shorter than orbital period \citep[e.g.,][]{Katz1973NPhS..246...87K, Barrett1988MNRAS, harvey1995superhumps, wood2009sph}. This phenomenon is thought to result from the retrograde precession of the tilted disks combined with the effects of mass stream from the secondary \citep[e.g.,][]{1985A&A...143..313B, patterson1999permanent, harvey1995superhumps}.

Our recent research has uncovered that the depth of eclipses, the brightness minima during eclipses, the amplitude of NSHs, and their frequencies all display periodic variations that align with the precession period of the tilted disks (e.g., TV Col, SDSS J0812 and HS 2325+8205 ; \citealp{Sun2024ApJ,Sun2023arXiv,2024arXiv240704913S}). These findings provide strong evidence for the existence of tilted disks and the origin of NSHs associated with the precession of tilted disks. Furthermore, we have observed a correlation between DN outbursts and the formation of NSHs, offering new perspectives on the origins of NSHs and the mechanisms behind DN outbursts (e.g., AH Her, ASAS J1420, TZ Per, and V392 Hya; \citealp{2023arXiv230905891S, 2023arXiv230911033S}).

Karachurin 12 was discovered by Raul Karachurin, and we used the naming convention of the American Association of Variable Star Observers (AAVSO). This source is also known as 
FBS1726+618, ASASSN-V J172710.78+614528.0, and SDSS J172710.79+614527.8. Kato et al. (2018)(vsnet-chat 7938)\footnote{http://ooruri.kusastro.kyoto-u.ac.jp/mailarchive/vsnet-chat/7938} suggests that Karachurin 12 be classified as a Z Cam-type DN.
There has been no in-depth analysis of Karachurin 12 to date, making this paper the first detailed study. This paper will determine the parameters and light curve characteristics, and use dip as an index and NSH as a probe to investigate the IW And-type phenomenon in Karachurin 12, offering new observational evidence to better understand this phenomenon.

The structure of this paper is outlined as follows:
Section \ref{sec:style} details the data sources utilized in this paper.
Section \ref{sec:Period} focuses on identifying periodic information for Karachurin 12.
Section \ref{sec:IWAnd} provides an in-depth analysis of the IW And-type phenomenon.
Section \ref{sec:NSHIWAnd} explores the evolution of the NSH in relation to the IW And-type phenomenon.
Section \ref{sec:DISCUSSION} discusses the physical processes of the IW And-type phenomenon in the context of current research advances and the results presented in this paper.
Finally, Section \ref{sec:CONCLUSIONS} presents the conclusions.

\begin{figure}
\centerline{\includegraphics[width=0.6\columnwidth]{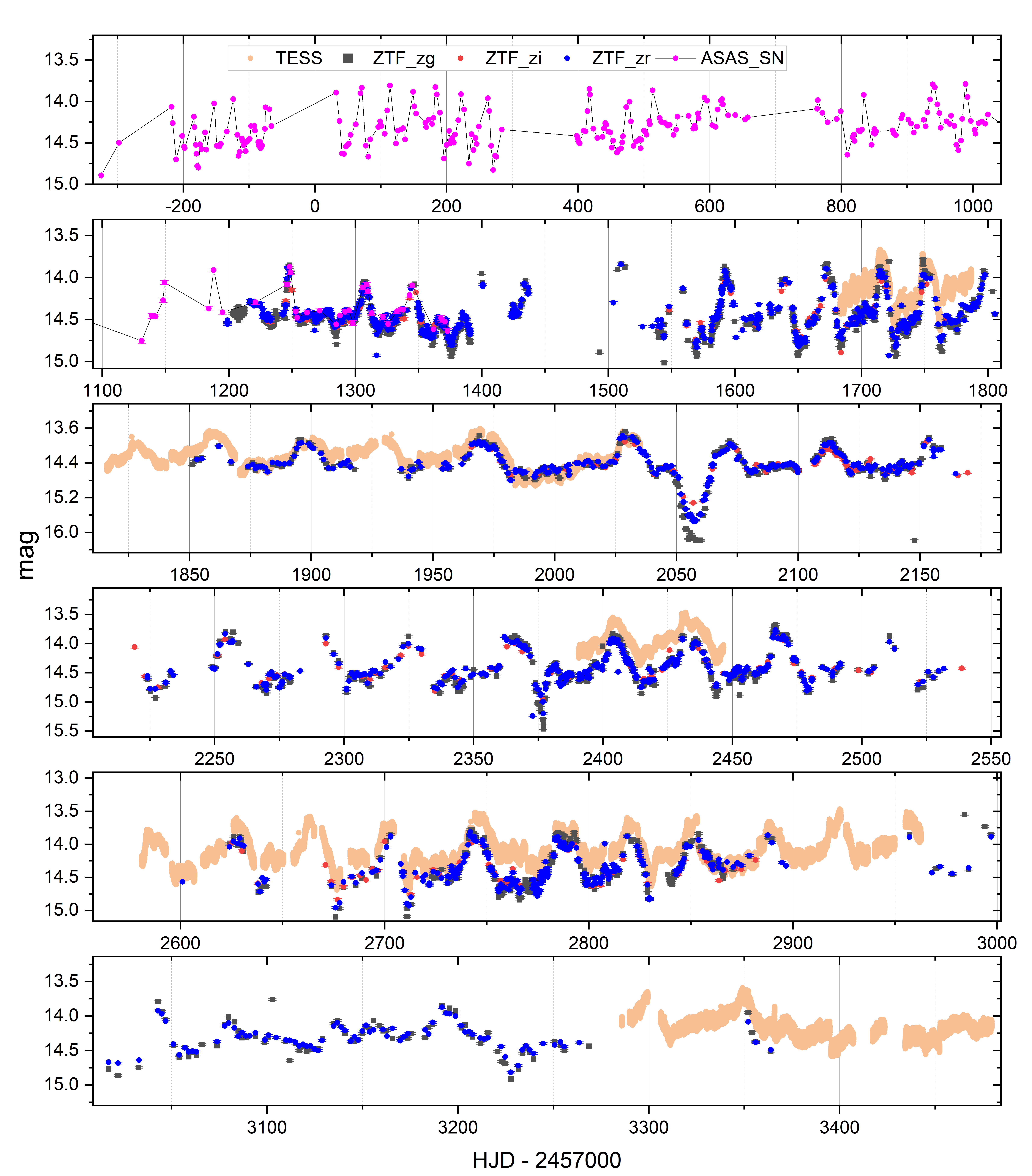}}
\caption{The light curve for Karachurin 12 is constructed from observations by ASAS-SN, ZTF, and TESS. The different symbols in the plot represent the various observational datasets, as indicated in the top panels. TESS observations are initially recorded in barycentric Julian date, but for comparison purposes, they have been converted to heliocentric Julian date.\label{light}}
\end{figure}

%%%%%%%%%%%%%%%%%%%%%%%%%%%%%%%%%%%%%%%%%%%%%%%%%%%%%%%%%%%%%%%%%%%%%%%

\section{Sources of Observation Data} \label{sec:style}

The All-Sky Automated Survey for Supernovae (ASAS-SN; \citealp{2014ApJ...788...48S}) is a long-term initiative dedicated to rapid supernova monitoring across the sky (V < 17 mag) and also tracking numerous variable stars \citep{2019MNRAS.486.1907J}. Karachurin 12 was photometrically observed in the V-band by ASAS-SN from HJD 2456675.131 to HJD 2458372.751, with an average magnitude of 14.33 mag (see Fig. \ref{light}). Data were retrieved from the ASAS-SN Variable Stars Database \footnote{https://asas-sn.osu.edu/variables}.

The Zwicky Transient Facility (ZTF; \citealp{2019ZTF}) is a northern-sky optical survey focused on high-cadence time-domain astronomy, utilizing the 48-inch Samuel Oschin Schmidt Telescope at Palomar Observatory. ZTF operates with three custom filters: ZTF$ \_ $zg, ZTF$ \_ $zr, and ZTF$ \_ $zi. Karachurin 12 was observed in these bands over approximately 6 years (from HJD 2458198.8895 to HJD 2460363.9715; see Fig. \ref{light}). Data were obtained from the Lasair database \footnote{https://lasair.roe.ac.uk/}.

The Transiting Exoplanet Survey Satellite (TESS; \citealp{Ricker2015journal}) is primarily designed for exoplanet detection but has also collected a significant amount of variable star data. TESS conducts sky surveys in sectors, observing each sector for approximately one month in the 600 to 1000 nm wavelength range, utilizing both long-cadence (30 minutes) and short-cadence (2 minutes) modes. The Mikulski Archive for Space Telescopes (MAST) provides light curves from both Simple Aperture Photometry (SAP) and Presearch Data Conditioning Simple Aperture Photometry (PDCSAP). While PDCSAP aims to calibrate discontinuities in observed data across different sectors and remove systematic errors, excess flux, and other anomalies (see \citealp{2010SPIE.7740E..1UT} and \citealp{2012PASP..124..963K} for details), this process inadvertently distorts the IW And-type phenomenon, as PDCSAP removes the long term trends. This paper employs SAP data and excludes the data points removed by PDCSAP. Although using SAP data inevitably introduces discontinuities from different sectors, as evidenced by the fluctuations of the TESS curves around the ZTF data shown in Figure \ref{light}, it remains more advantageous for this study than PDCSAP. Karachurin 12 was observed across 35 sectors over approximately 5 years in short-cadence mode (see Fig. \ref{light}), with detailed observation data listed in Table \ref{tessLOG}. TESS data were downloaded from the MAST\footnote{https://mast.stsci.edu/}.

\begin{figure}
	\gridline{\fig{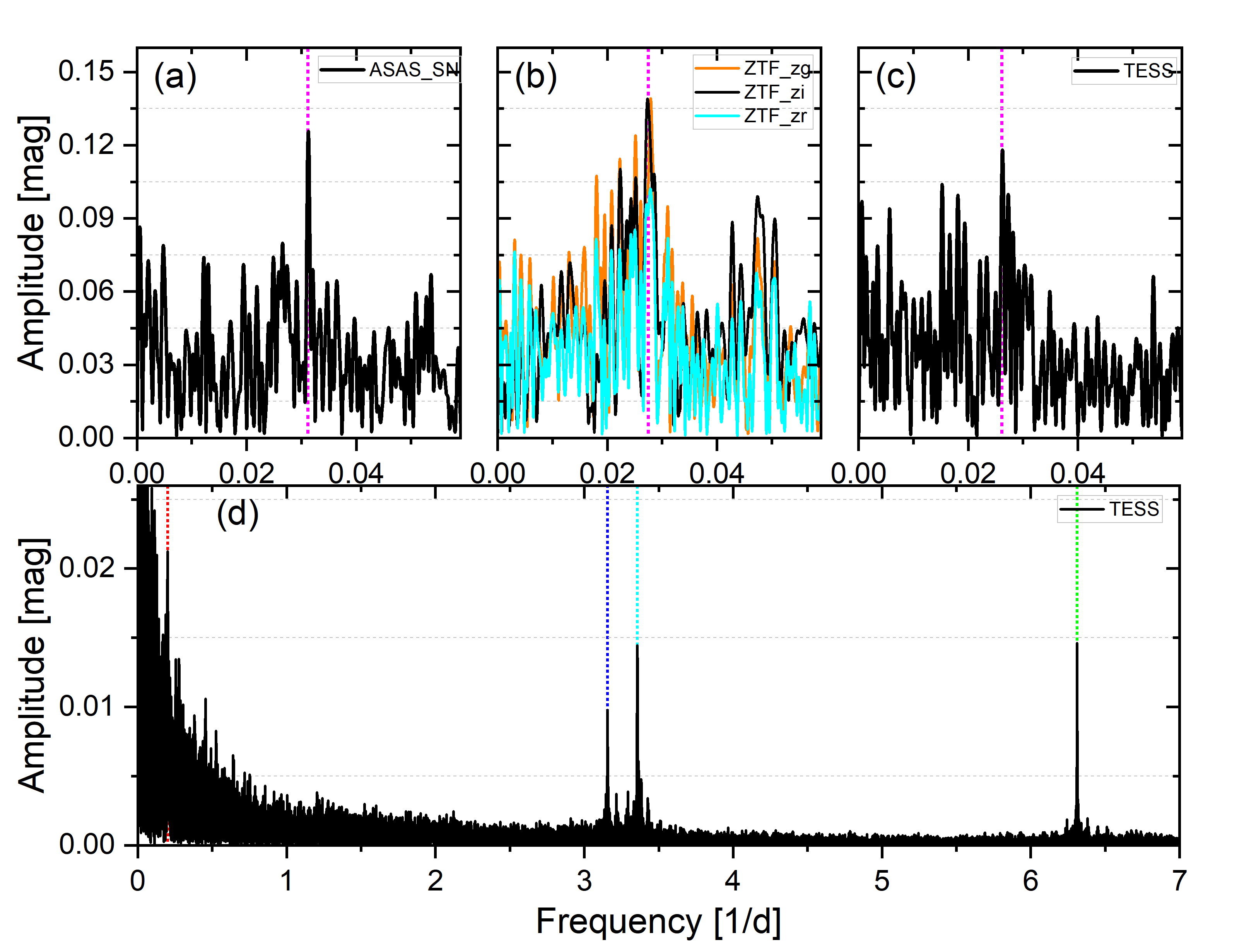}{0.5\textwidth}{}
		\fig{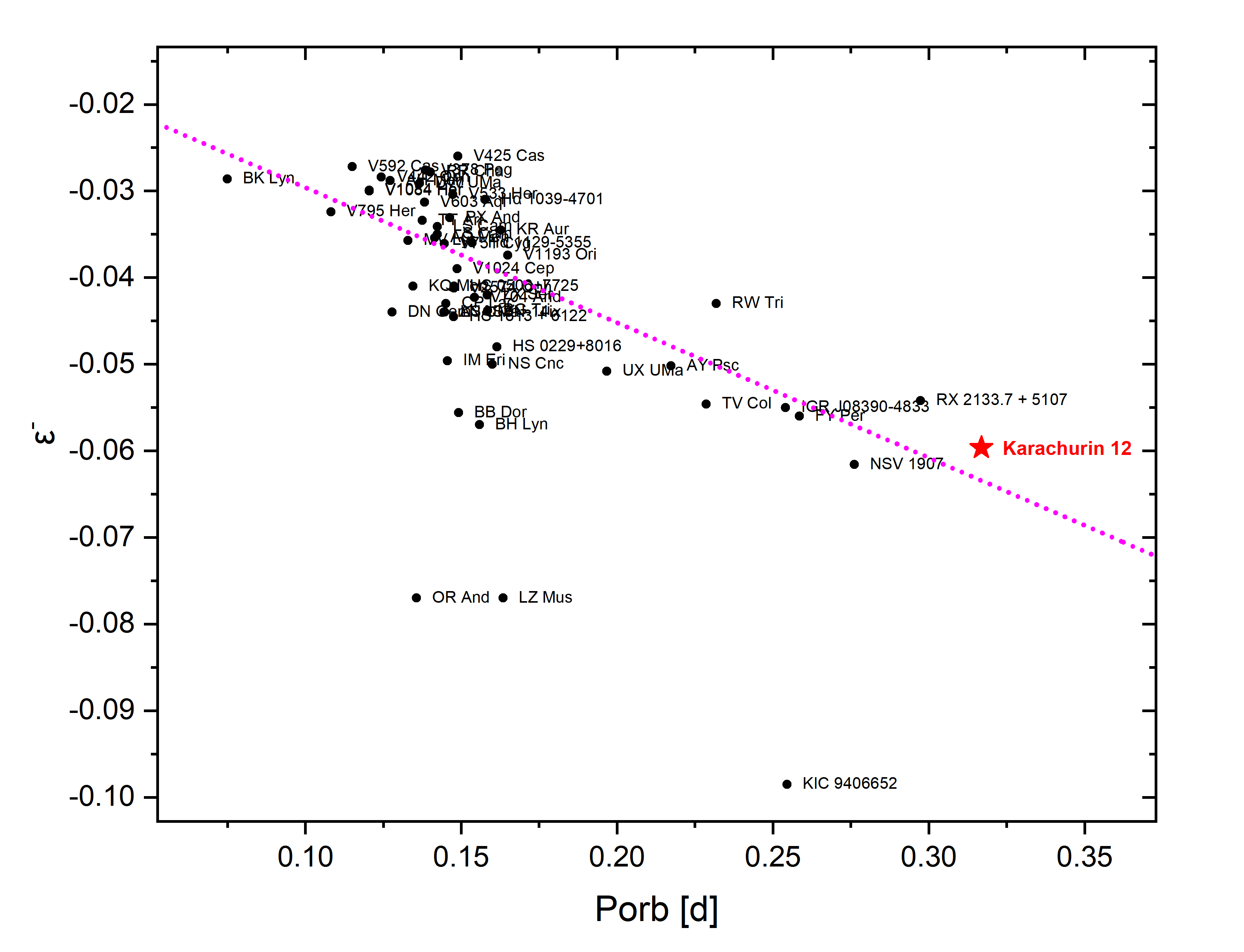}{0.5\textwidth}{}
	}
	\caption{Frequency spectrograms of Karachurin 12 and $\epsilon^{-}$ as a function of orbital period. (a) Frequency analysis for ASAS-SN data; (b) Frequency analysis for ZTF data with color-coded bands; (c) Frequency analysis of TESS photometry highlighting low-frequency components; (d) Same data as panel (c) with emphasis on small amplitude and high-frequency signals. Vertical dashed lines (magenta, red, blue, cyan, green) denote outburst signals, accretion disk precession, orbital frequency, NSH frequency, and the second harmonic of the orbital frequency. Black data points and magenta dotted lines in the right panel are from \cite{BruchIII}, with the red pentagram marking Karachurin 12.\label{fft}}
\end{figure}

%%%%%%%%%%%%%%%%%%%%%%%%%%%%%%%%%%%%Analysis%%%%%%%%%%%%%%%%%%%%%%%%%%%%%%%%%%%%%%%%

\section{Period determination} \label{sec:Period}

The ASAS-SN, ZTF, and TESS observations all exhibit IW And-type phenomena with notable variations in outburst amplitudes. We analyzed the periods for each dataset individually using the \texttt{Period04} software (details in \citealp{Period04}). For ZTF, frequency analyses were conducted separately on the ZTF$ \_ $zg, ZTF$ \_ $zr, and ZTF$ \_ $zi bands. The period of IW And-type phenomena was determined to be 31.98(5) days from ASAS-SN data (see Fig. \ref{fft} a) and 38.095(2) days from TESS data (see Fig. \ref{fft} c). For ZTF, the cycle periods were 35.82(3) days for ZTF$ \_ $zg, 35.97(3) days for ZTF$ \_ $zr, and 36.60(6) days for ZTF$ \_ $zi (see Fig. \ref{fft} b). Averaging these values, the period of the IW And-type phenomena is found to be 35.69(3) days.

The TESS data, with a 120-second exposure time, provides crucial insights into detailed variations in our study. Frequency analysis reveals periodic signals with periods of f$_{2}$ = 0.201662(6) d$^{-1}$, f$_{3}$ = 3.155674(13) d$^{-1}$, f$_{4}$ = 3.355861(9) d$^{-1}$, and f$_{5}$ = 6.311376(7) d$^{-1}$, in addition to the IW And-type phenomenon. 
Notably, f$_{5}$ is approximately 2 × f$_{3}$, confirming it as the second harmonic of f$_{2}$. The relationship among f$_{2}$, f$_{3}$, and f$_{4}$ can be expressed as f$_{2}$ $ \approx $ f$_{4}$ - f$_{3}$, reflecting the connection between the precession period of the accretion disks, the orbital period, and the NSH period (1/P$_{\text{prec}}$ = 1/P$_{\text{nsh}}$ - 1/P$_{\text{orb}}$). 
From this, we infer that f$_{2}$ corresponds to a precession signal with a period of 4.9588(2) days for the tilted disks, f$_{3}$ represents an orbital period of 0.3168895(13) days, f$_{4}$ corresponds to the NSH period of 0.2979861(8) days, and f$_{5}$ denotes the second harmonic of the orbital period. 
It should be noted that f$_{2}$ is not strictly equal to 1/P$_{\text{nsh}}$ - 1/P$_{\text{orb}}$ (0.200187(22) d$^{-1}$), and the observed difference of 0.001475(28) d$^{-1}$ is outside the error range. This discrepancy may arise from the data processing, but it could also reflect a real change in the tilted disks. We suggest that this difference remains valuable for the study of accretion disk precession.
No eclipses were observed in Karachurin 12, aside from the ellipsoidal modulation, suggesting that it is a low-inclination CV.
Excess is a key parameter in the study of NSHs, defined as $\rm \epsilon^{-} = (P_{\text{nsh}} - P_{\text{orb}}) / P_{\text{orb}}$. For Karachurin 12, $\epsilon^{-}$ is determined to be -0.059653(7). Comparing this value with the empirical relationship between orbital period and $\epsilon^{-}$ derived by \cite{BruchIII} from a large sample of NSHs, we find a general agreement with their results (see Fig. \ref{fft}).

\begin{figure}
		\centering
	\gridline{\fig{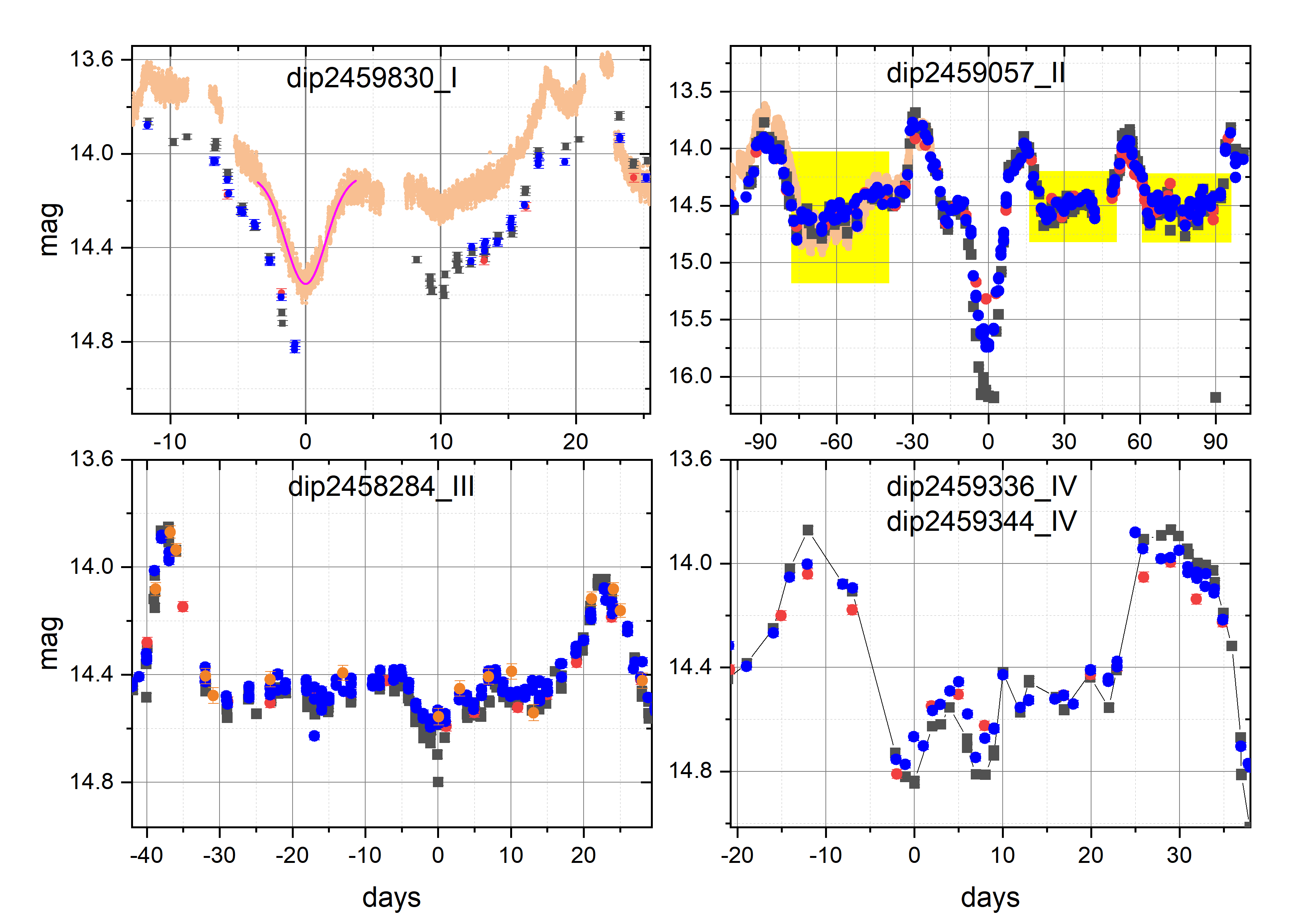}{0.52\textwidth}{}
		\fig{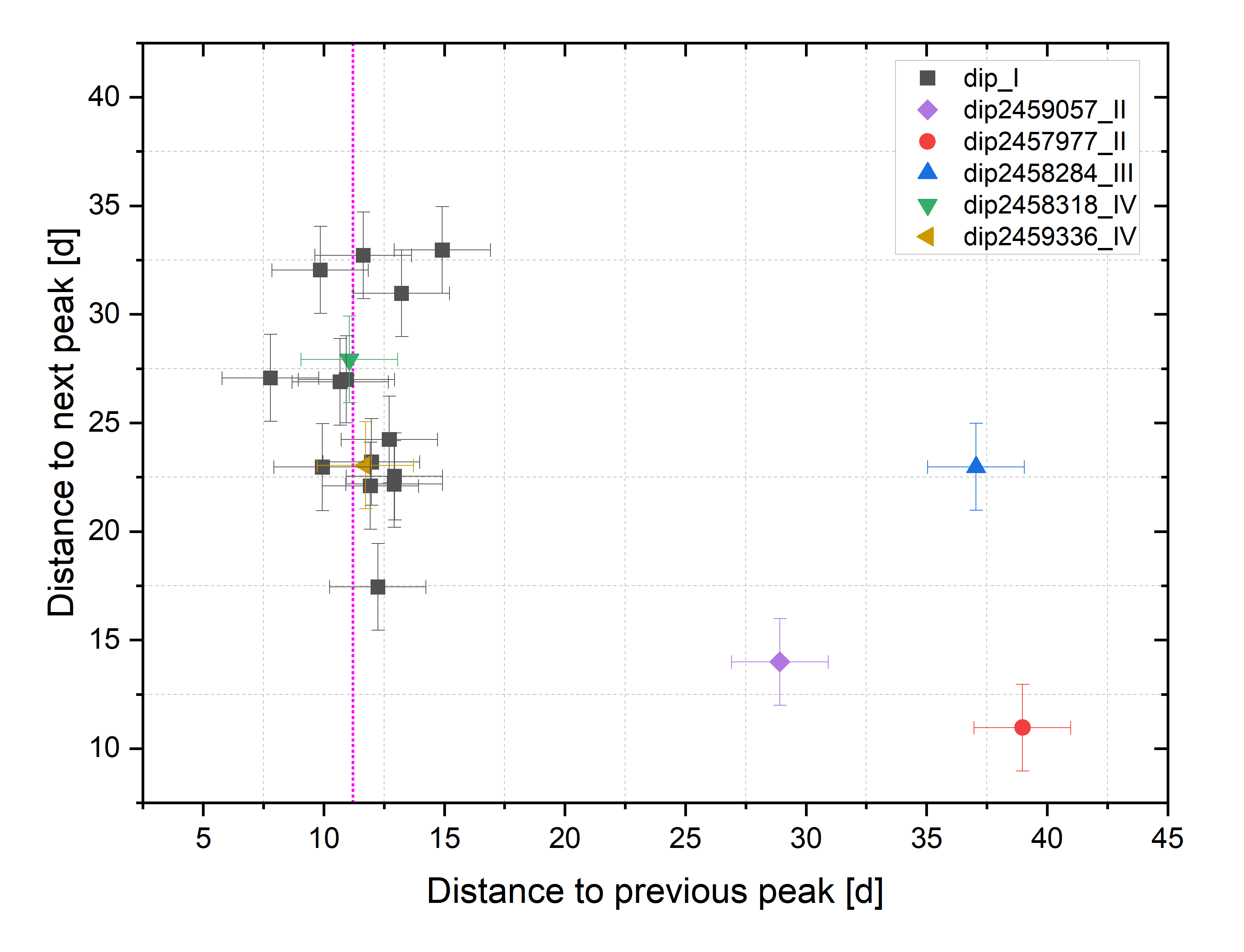}{0.48\textwidth}{}
	}
	\caption{Different types of dips and cycles for Karachurin 12. Left panel: The magenta curve in dip2459830 is a Gaussian fit; the yellow region in dip2459057 has no significant dip, the data points color code is the same as in Figure \ref{light}. Right panel: Statistical correlation between the time intervals from the dip to the preceding outburst peak and to the following outburst peak. \label{dip}}
\end{figure}

\section{IW And-type phenomenon} \label{sec:IWAnd}

\subsection{Identification of dips} 

Typical IW And objects are characterized by standstill or quasi-standstill phases interrupted by brightening events or outbursts, which are collectively referred to as outbursts. Unlike Z Cam-type stars that return to a quiescent state, these systems undergo damping oscillations following outbursts, resulting in a cycle of standstill–outburst–damping oscillations \citep[e.g.,][]{Kato2019PASJ, 2020PASJIMEri, 2021ApJHOPuppis}.

In the case of Karachurin 12, we observe a notable IW And-type phenomenon where the typical cyclic sequence includes a quasi-standstill phase interrupted by an outburst, followed immediately by a dip. Unlike other IW And objects, the quasi-standstill phase in Karachurin 12 does not show significant damping oscillations. Instead, it reflects the signal of accretion disk precession, although this precession signal is not prominent in every quasi-standstill phase (see Fig. \ref{16dip}). In addition to the standard cycles, Karachurin 12 exhibits unique variations, with relatively stable outburst and quasi-standstill phases, and variations mainly arising from the dip. Therefore, we use the dip as an index to study these special cycles.

The IW And-type phenomenon has been documented through surveys conducted by ASAS-SN, ZTF, and TESS. ASAS-SN data, being relatively dispersed, contrasts with the more precise photometric results provided by TESS. Our initial approach involved identifying the positions of the dips based on their occurrence times and labeling them accordingly (e.g., dip2459830, as shown in Table \ref{diptab}). To determine the precise location of these dips, we employed Gaussian fitting and identified the minimum points of the dips.

For TESS data, where the dip profiles were relatively complete, we used Gaussian fitting to calculate the dip parameters. For ZTF and ASAS-SN data, we determined the timing of the dips from the minimum points. For ZTF data, the ZTF$ \_ $zg band, with its greater detail and larger amplitude variations, was used to define the dip parameters. In the ASAS-SN dataset, only one particularly complete profile was identified. In total, we recognized 23 dips, 8 of which were determined through Gaussian fitting (see Table \ref{diptab}).

\subsection{Different types of dip and corresponding cycles} 

Next, we use dips as indices to study the cyclical nature of the IW And-type phenomenon. Among the identified dips, 16 are classified as normal cycles, characterized by a quasi-standstill phase interrupted by an outburst, followed by a dip. We designate these as Type$ \_ $I dips (see Fig. \ref{16dip}). For Type$ \_ $I, eight dips were analyzed using Gaussian fitting (see Figs. \ref{16dip} and \ref{dip}). The width of each dip was determined as the full width at half maximum (FWHM) of the fit, with an average width of 3.86(10) days and a depth of 0.32(8) mag. The quasi-standstill duration typically ranges from about 14.0 to 14.5 mag and lasts between 10 and 20 days.

The deepest dip observed in Karachurin 12 is dip2459057 (see Fig. \ref{dip}), with a depth of approximately 1.5 mag and a duration of about 15 days. This dip is notably 4 to 5 times wider and deeper than a typical dip. Unusually, it was discovered not following an outburst phase but rather within a cycle that included an outburst, a quasi-standstill, and then the dip (outburst - quasi-standstill - dip - outburst). This deviation from the Type$ \_ $I pattern leads us to classify it as Type$ \_ $II. Another example of Type$ \_ $II is dip2457977 (see Fig. \ref{fig:6dip}).

Another distinct case is dip2458284 (see Fig. \ref{dip}), where the cycle consists of an outburst followed by a 30-day standstill, which is then truncated by a dip. This is followed by a new standstill phase, truncated by another outburst, resulting in the cycle: outburst - quasi-standstill - dip - quasi-standstill - outburst. We designate this as Type$ \_ $III. The key difference between Type$ \_ $III and Type$ \_ $II is that Type$ \_ $III is followed by another standstill. Similar variations may be present in dip2458375 (see Fig. \ref{fig:6dip}), though the data is less clear due to gaps.

In dips 2459336 and 2459344, two dips occur within a single IW And cycle, approximately 8 days apart (see Fig. \ref{dip}). This pattern can be attributed to precession (4.9588(2) days) followed by a quasi-standstill. A similar pattern is observed in dips 2458318 and 2458326 (see Fig. \ref{fig:6dip}). Unlike Type$ \_ $I, where only a single dip is observed, these double dips are classified as Type$ \_ $IV. This type of variation may be analogous to the damping oscillations observed in IM Eri \citep{2020PASJIMEri}. However, further observational data is needed to confirm these double-dip occurrences.

To compare the four types of dips, we calculated the time intervals between clearly defined dips and the preceding and following outburst peaks, with an uncertainty of 2 days. Results show that Type$ \_ $I dips typically occur about 11 days after the previous outburst peak and 15 to 35 days before the next outburst peak (see Fig. \ref{dip}). This suggests that the interval between Type$ \_ $I dips and the preceding outburst is relatively stable, while the duration of the quasi-standstill phase varies.
In contrast, Type$ \_ $II dips occur much further from the preceding outburst and closer to the subsequent one compared to Type$ \_ $I (see Fig. \ref{dip}). Type$ \_ $III dips are separated from both preceding and following outbursts by more than 20 days (see Fig. \ref{dip}). For Type$ \_ $IV, the first dip in the double-dip structure shows intervals to the preceding and following outbursts consistent with those of Type$ \_ $I.

By indexing dips according to their position within the cycle, we observed various dip types and their corresponding cycles in Karachurin 12. Our study reveals the following:
(i) Among the 23 dips analyzed, 18 occur after an outburst, indicating that dips are more likely to follow an outburst phase.
(ii) However, a dip does not always follow an outburst phase, as seen in Type$ \_ $II and Type$ \_ $III (see Fig. \ref{dip}).
(iii) A dip does not always have to be followed by a quasi-standstill and can directly transition into an outburst phase, as observed in Type$ \_ $II (see Fig. \ref{dip}).
(iv) Not every cycle must include a dip (see the yellow regions of dip2459057 in Fig. \ref{dip}). While various dips and their corresponding cycles have been identified in Karachurin 12, further verification with additional stars and data is necessary.

\begin{figure}
	\centering
	\gridline{\fig{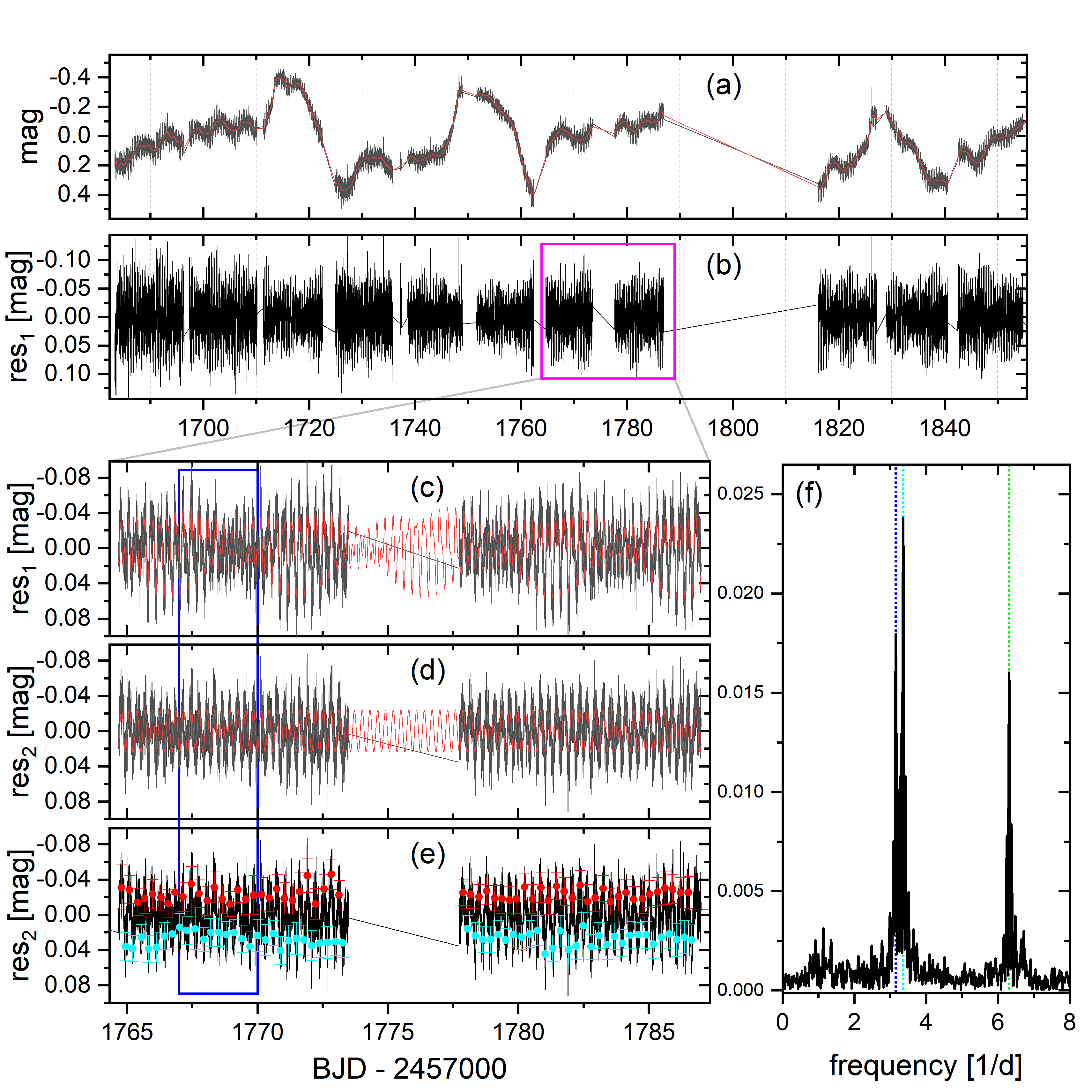}{0.5\textwidth}{}
		\fig{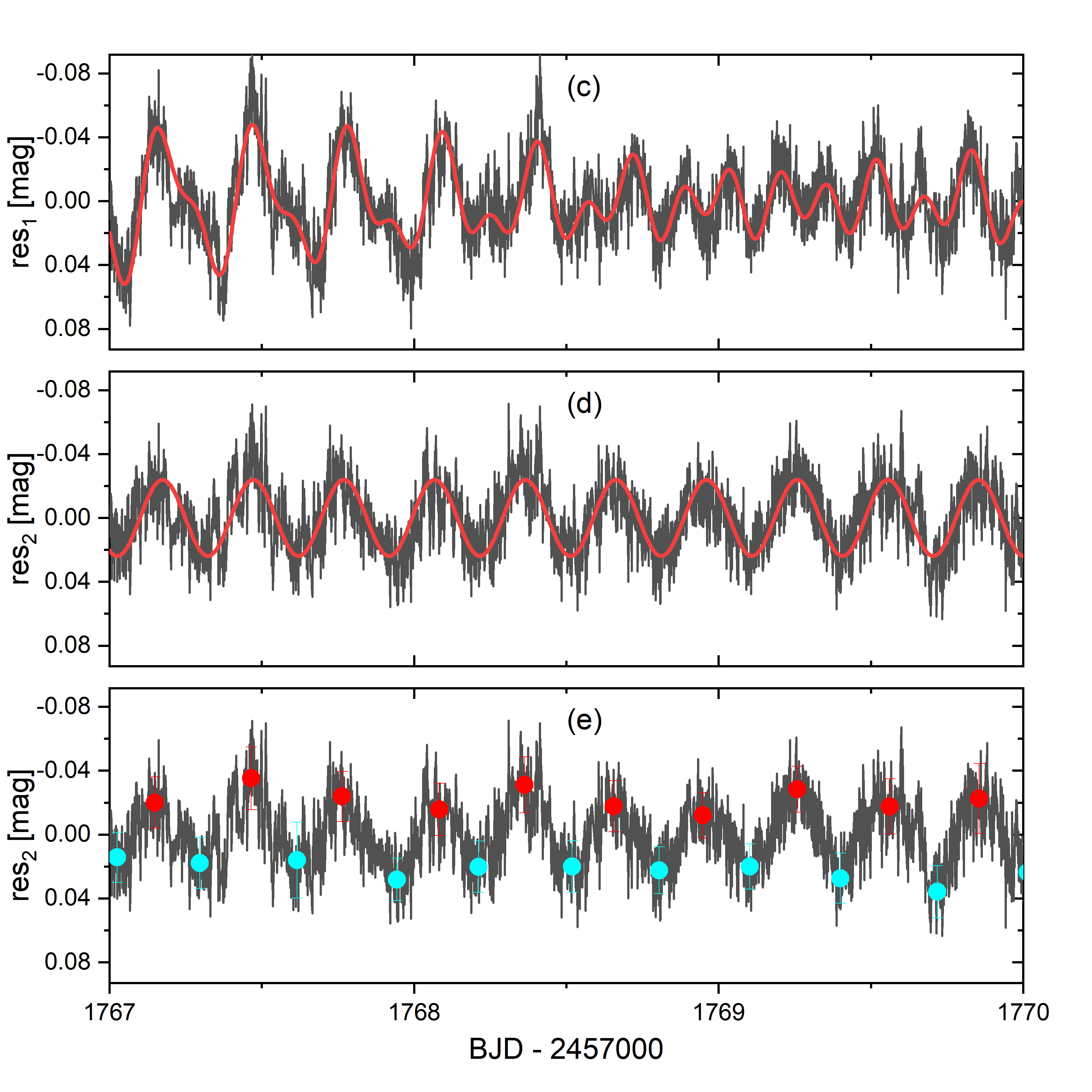}{0.5\textwidth}{}
	}
	\caption{Example of NSH analysis. Left panel: (a) TESS photometric data, with the red solid line representing the LOWESS fit used to remove long-term trends; (b) residuals from the LOWESS fit; (c) the magenta rectangle in panel (b) highlights the corresponding curve, with the red curve showing the fit for the orbital signal, NSH, and twice the orbital signal; (d) the light curve after removing the orbital signal, with the red solid line showing the sinusoidal fit corresponding to the NSH signal; (e) the black curve is consistent with panel (d), while the red and magenta markers represent the maxima and minima of the NSH obtained from Gaussian fitting; (f) the periodogram corresponding to the curve in panel (c), with blue, magenta, and green vertical lines representing the orbital signal, NSH, and twice the orbital harmonic, respectively. The right panel is an enlarged view of the blue rectangular box in the left panel.\label{Example}}
\end{figure}

\section{Evolution of NSH with the IW And-type phenomenon} \label{sec:NSHIWAnd}

\subsection{Extraction of NSH information} \label{sec:NSHinformation}

Current research indicates that the IW And-type phenomenon is associated with tilted accretion disks. To explore the relationship between NSHs and the IW And-type phenomenon and to provide new observational evidence, we propose using NSHs as a diagnostic tool. The origin of NSHs is closely linked with mass transfer streams from the secondary star and variations within the accretion disks. Given their higher completeness and continuity compared to precession signals observed in Karachurin 12, NSHs are an ideal probe for this investigation.

To study the evolution of NSHs, we follow the methodology outlined by \cite{2023arXiv230911033S} for excluding long-term trends and orbital signals. We employed locally weighted regression (LOWESS; \citealp{cleveland1979robust}) to remove the long-term trend (see Fig. \ref{Example}a). Orbital signals were subsequently removed using \texttt{Period04} software (details in \citealp{Period04}). The evolution of NSH amplitude and period is calculated as follows:

(i) Frequency Analysis: Using \texttt{Period04}, we performed a frequency analysis of the light curves after removing the long-term trend for each sector, identifying signals with signal-to-noise ratios greater than 3. The periodogram for each sector is shown in Figure \ref{frequency}, and the statistical results are listed in Table \ref{tab:A7}.

(ii) Amplitude and Phase Determination: The amplitude and phase information for each signal were obtained using the fitting equation provided by \texttt{Period04} (see Fig. \ref{Example}c):
\begin{equation} \label{eq:sine1}
	\text{mag}(t) = Z + \sum \text{Amplitude}_i \cdot \sin(2\pi \cdot (\text{frequency}_i \cdot t + \text{phase}_i))
\end{equation}
where $Z$, $\text{Amplitude}_i$, $\text{frequency}_i$, and $\text{phase}_i$ are the fitted intercept, amplitude, frequency, and phase, respectively.

(iii) Isolation of NSH Signal: Based on the fit results, we adjusted the detrended light curve to isolate the NSH signal by subtracting both the orbital signal and its second harmonic (as illustrated in Fig. \ref{Example}c). A sinusoidal fit was then applied to the residual data (see Fig. \ref{Example}d).

(iv) Calculation of Amplitude: To determine the NSH maxima and minima, we applied a Gaussian fit (see Fig. \ref{Example}e). The NSH amplitude was calculated using the following equation:
\begin{equation} \label{eq:am}
	\Delta \text{Amplitude}_{\text{nsh}} = \frac{\left( \text{minima}_{\text{before, mag}} + \text{minima}_{\text{after, mag}} \right) / 2 - \text{maxima}_{\text{mag}}}{2}
\end{equation}
Here, $\text{minima}_{\text{before, mag}}$ and $\text{minima}_{\text{after, mag}}$ refer to the minima before and after the NSH maxima, respectively. The calculated amplitudes are shown as black points in Figure \ref{cwt}b of the left panel. Note that some NSH amplitudes could not be determined due to significant weakening of the NSHs.

(v) Segmented Frequency Analysis: To validate the NSH results from step (iv) and address issues with unclear contours, we performed a segmented frequency analysis of the light curves after removing long-term trends and orbital signals. The data were initially divided into one-day segments for periodic analysis (indicated by the blue points in Fig. \ref{cwt}b of the left panel and Fig. \ref{fig:cwt2}c). The data were then re-divided into two-day segments for further analysis (shown by the red points in Figs. \ref{cwt}b of the left panel and Fig. \ref{fig:cwt2}c). This approach aimed to provide a clearer profile of the frequency and amplitude evolution of the NSHs.

(vi) Continuous Wavelet Transform (CWT): Finally, in agreement with \cite{Sun2023MNRAS,2023arXiv230911033S}, we applied the CWT \citep{polikar1996wavelet} to the light curves after removing long-term trends and orbital signals. This method complemented steps (iv) and (v) and facilitated cross-validation of the NSH information (see Fig. \ref{cwt}c in the left panel). In the two-dimensional (2D) power spectrum of the CWT, different colour stripes indicate the relative intensity of the NSHs, with redder stripes indicating larger NSH amplitudes. Figure \ref{cwt}c in the left panel exhibits the NSH amplitude variation with outbursts in general agreement with the results obtained in steps (v) and (iv).

\subsection{The evolution of NSH} 

\begin{figure}
		\centering
\gridline{\fig{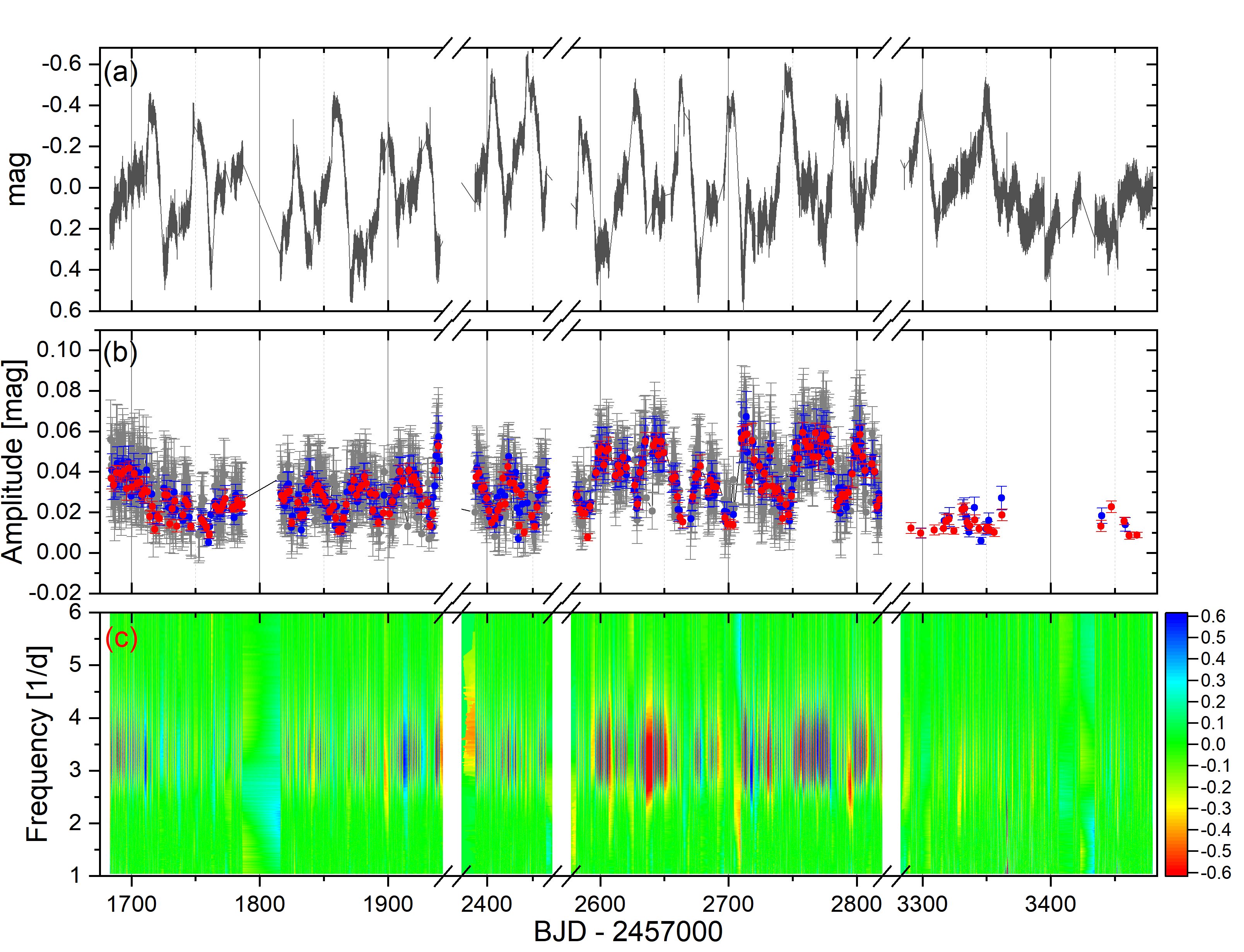}{0.47\textwidth}{}
	\fig{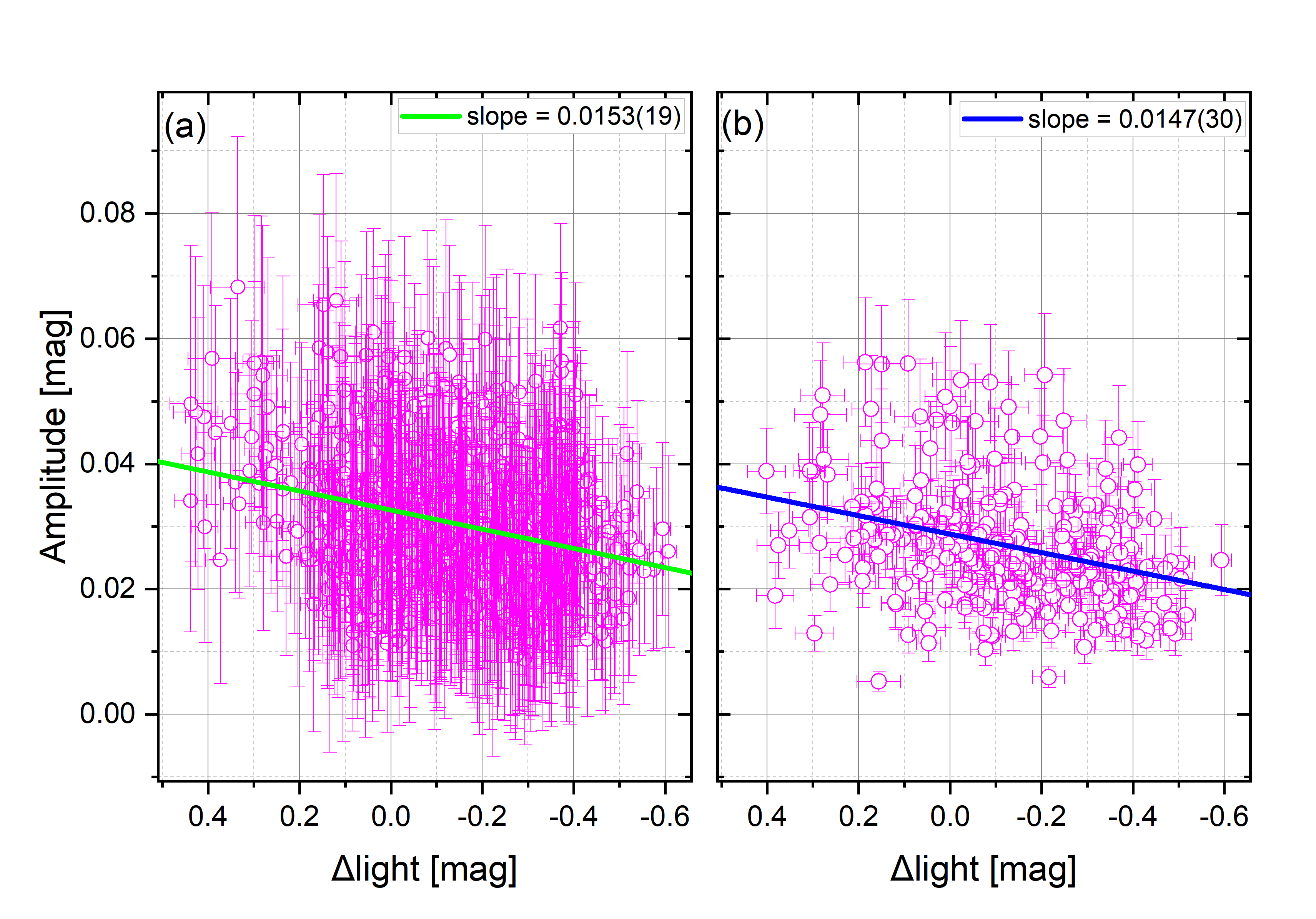}{0.53\textwidth}{}
}	
	
	\caption{Comparison of NSH amplitude variations with the light curve. Left panel: (a) All TESS photometric data; (b) NSH amplitude variation curves, with black points representing half the difference between maxima and minima, and blue and red points corresponding to NSH amplitude results from segment frequency analyses with segments of 1 day and 2 days, respectively; (c) 2D power spectrum obtained using CWT for light curves after the long-term trend and orbital signals have been removed, with the colour stripes representing the relative intensity of the NSHs in the -0.6 to 0.6 scales. Redder colours indicate larger NSH amplitudes. The right panel shows the statistical relationship between the magnitude of the light curve and the NSH amplitude during the outburst. Panels (a) and (b) show the NSH amplitude calculated from the difference between the maxima and minima, and the segmented Fourier analysis at one-day intervals, respectively; the lines in the plots are linear fits. \label{cwt}}
\end{figure}

\begin{figure}[htbp]
	\centering
	\gridline{\fig{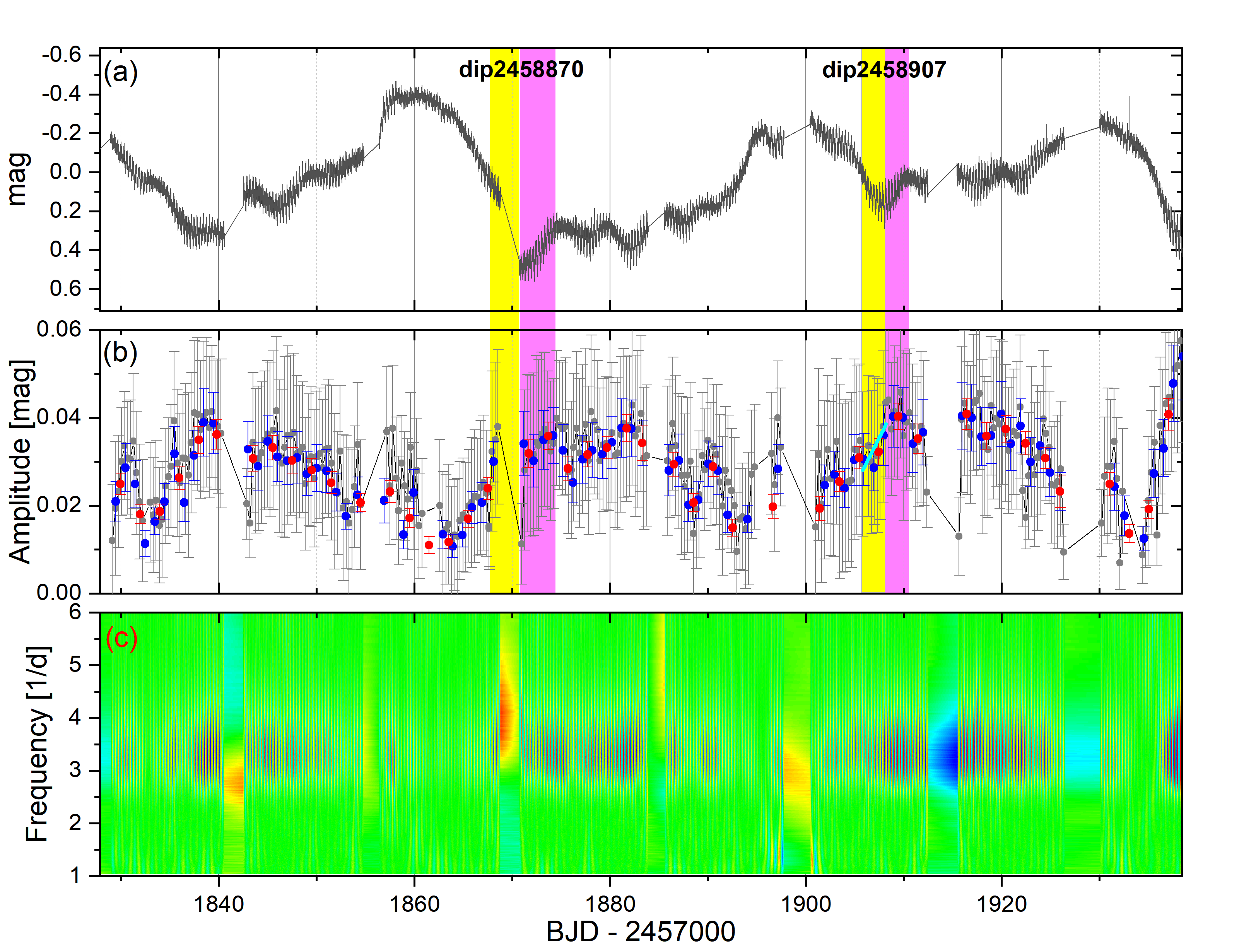}{0.5\textwidth}{(a)}
		\fig{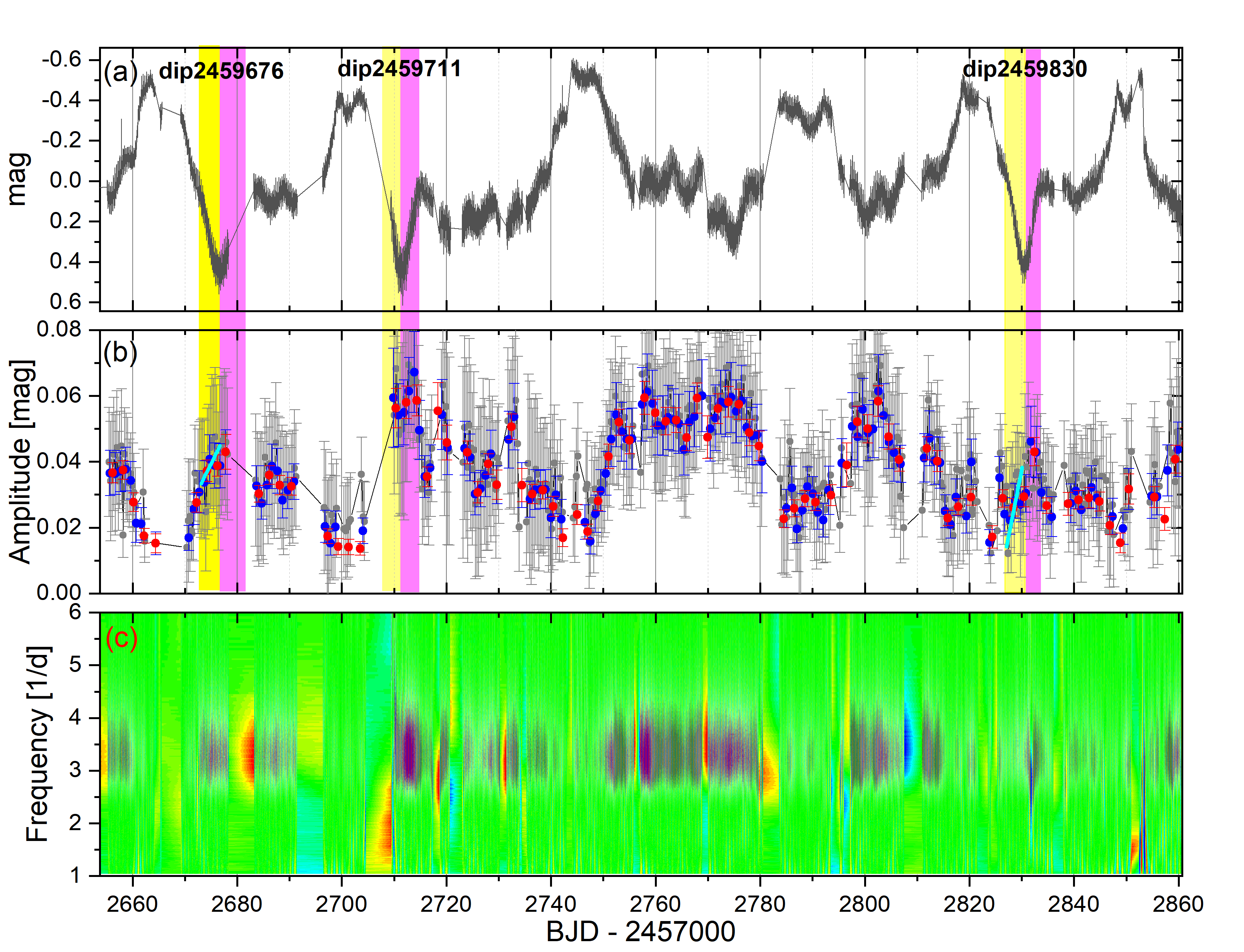}{0.5\textwidth}{(b)}
	}
	\gridline{\fig{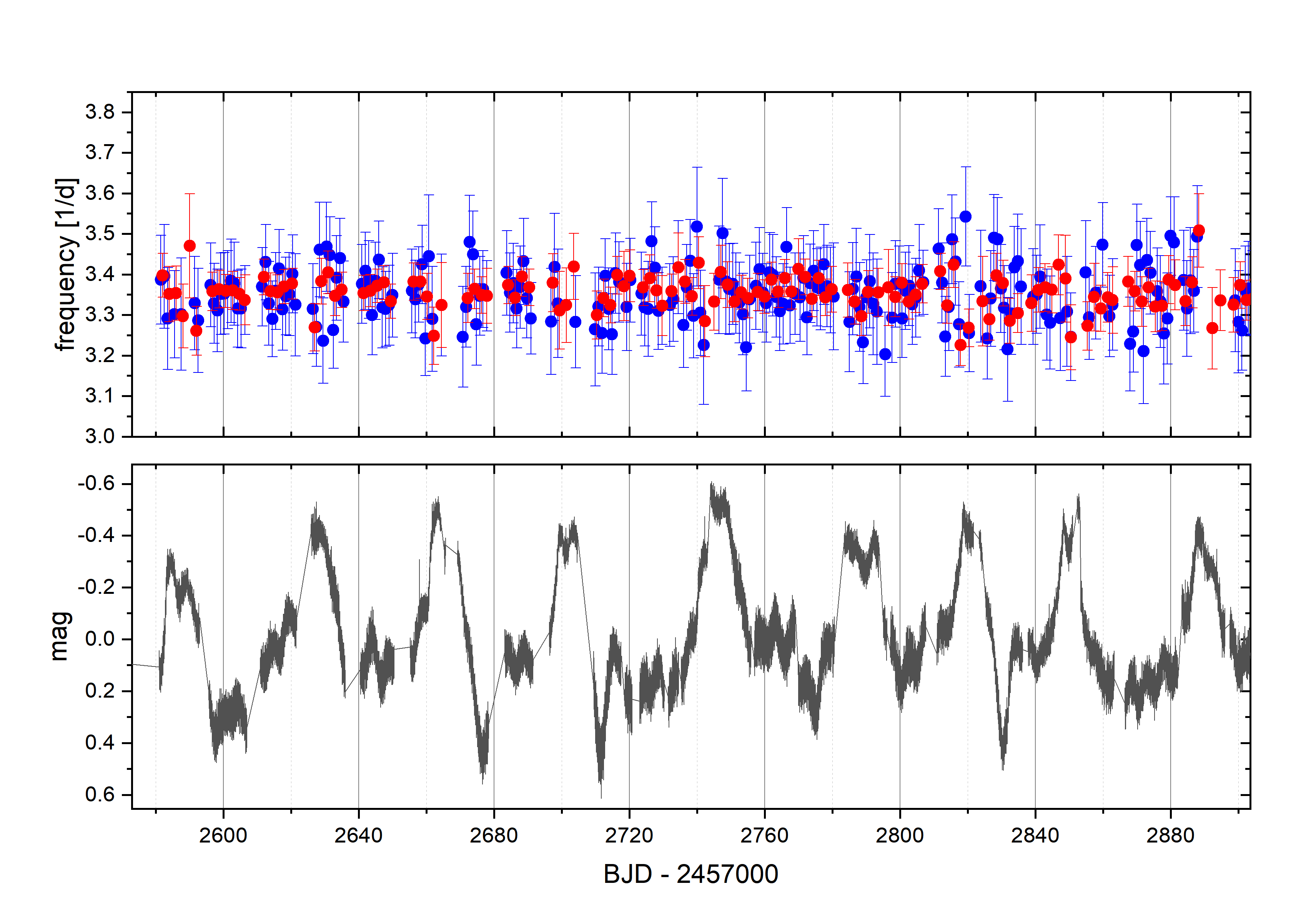}{0.5\textwidth}{(c)}
		\fig{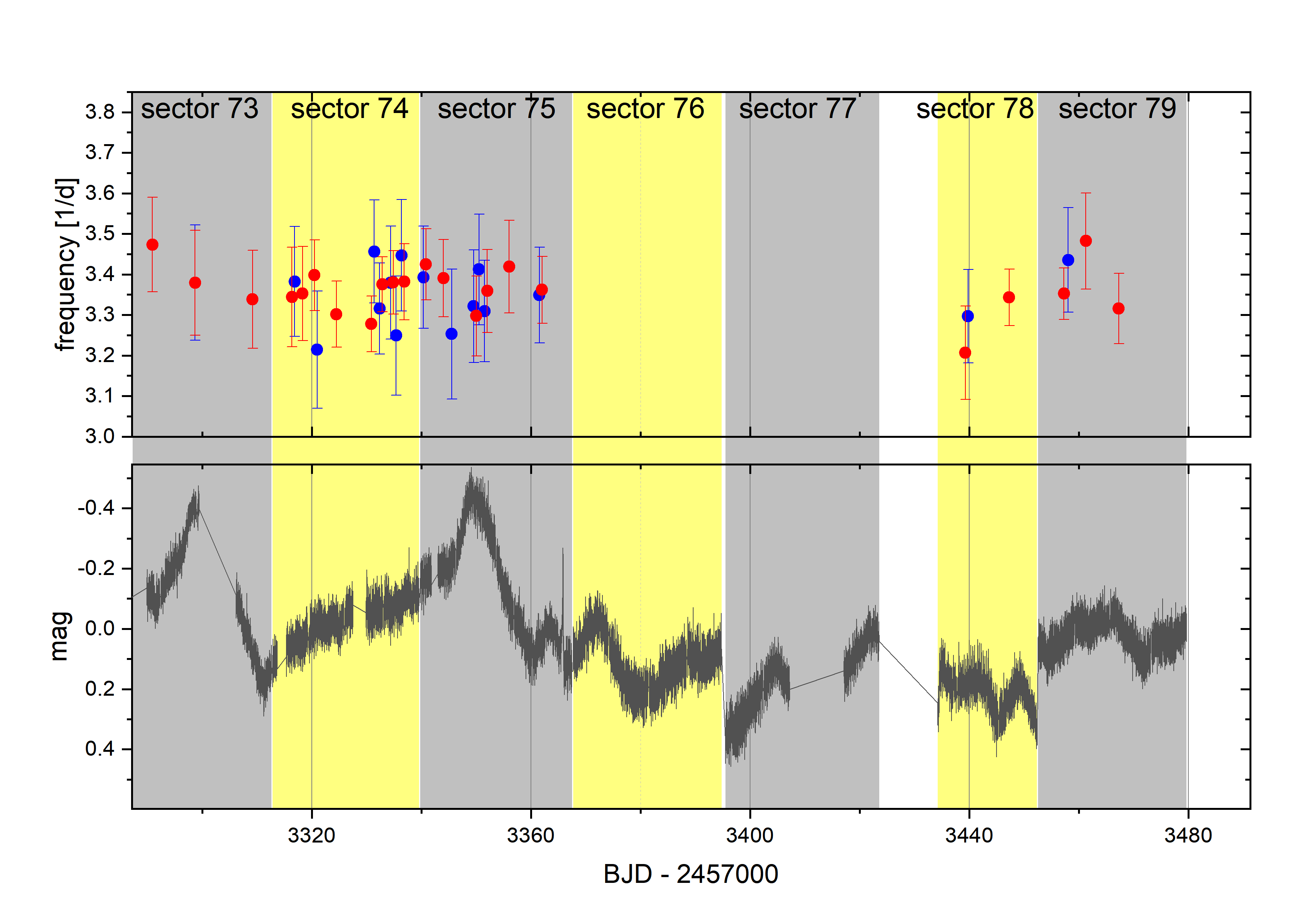}{0.5\textwidth}{(d)}
	}
	\caption{Evolution of the NSH amplitude and frequency with the light curve. (a) and (b) are from a local zoom of Fig. \ref{cwt} in the left panel, where the yellow area corresponds to the ingress dip, and the magenta area is the egress dip, and the fluorescent curve is a linear fit;
		(c): Results of the segmental frequency analysis for the red and blue dots in the top panel, analyzed in segments of 2 and 1 days;
		(d): Consistent with panel (c), but primarily displaying data from sectors 73 to 79, where the observations in sector 78 cover only about 18 days. }
	\label{fig:cwt2}
\end{figure}

A comparison of the results from steps (iv), (v), and (vi) reveals consistent trends with the following characteristics:

(a) The NSHs show a marked weakening starting from sector 73, becoming undetectable in sectors 76 and 77, and then weakly recovering in sectors 78 and 79. For detailed information, refer to Figures \ref{cwt}c in the left panel. Notably, no significant IW And-type phenomenon is observed after the outburst ends in sector 75 (see Fig. \ref{fig:cwt2}d), suggesting a potential link between the IW And-type phenomenon and NSHs.

(b) No discernible regularity in the period of NSHs was detected (see Fig. \ref{fig:cwt2}c), indicating that the NSH frequency does not vary with the IW And cycle in Karachurin 12.

(c) The amplitude of the NSH is correlated with the outburst phases, decreasing during the rise of an outburst and increasing during its recession. The maximum NSH amplitude is observed during the quasi-standstill phase (see Figs. \ref{cwt} and \ref{fig:cwt2}). Additionally, we selected the NSH amplitudes and the non-detrended light curves corresponding to the outburst phases for statistical analysis. The first analysis utilized the amplitude information obtained in step (iv) of Section \ref{sec:NSHinformation}, with the corresponding light curve representing the average magnitude between the two minima before and after the maxima.

The second calculation employed the NSH amplitudes derived from the one-day segmented Fourier analysis in step (v) of Section \ref{sec:NSHinformation}, with the average magnitude corresponding to the time spans of the light curve segments.
Linear fitting of the two statistical relationships yielded slopes of 0.0153(19) and 0.0147(30) (see Fig. \ref{cwt} in the right panel), respectively, further confirming a positive correlation between NSH amplitudes and magnitudes, and an inverse correlation with brightness.

(d) We analyzed the dip in three phases: ingress dip, minima, and egress dip. The ingress and egress dips are highlighted in yellow and magenta, respectively, in Figures \ref{fig:cwt2}a and \ref{fig:cwt2}b. A linear fit was applied to the NSH amplitude during the relatively complete ingress dip (fitting the black points). The results indicate that the NSH amplitude for the ingress dip continues to rise, reflecting the trend of the outburst recession phase. There is no evidence that the NSH amplitude evolution reverses at the dip minima, suggesting that the dip minima may not represent a turning point. Additionally, the NSH amplitudes during the egress dips of dip2458870 and dip2459711 continue to show an increasing trend.

%%%%%%%%%%%%%%%%%%%%%%%%%%%%%%%%%%%%DISCUSSION%%%%%%%%%%%%%%%%%%%%%%%%%%%%%%%%%%%%%%%%
\section{DISCUSSION} \label{sec:DISCUSSION}

The DIM explains the standstill phase observed in regular Z Cam systems by proposing that Z Cam has a mass transfer rate that approaches a critical value \citep{dubus2018testing}. Fluctuations in this mass transfer rate can drive the system into a hot state, resulting in a brightness standstill \citep[e.g.,][]{1983Meyer,1985MNRASLin,lasota2001disc}. According to the DIM, such a standstill should eventually end with a return to quiescence. However, the IW And-type phenomenon, where the standstill ends with an outburst, poses a challenge to this model. The DIM alone cannot account for recurring outbursts while the accretion disks remains in a hot state \citep{2020AdSpR..66.1004H}.

The origin of NSHs is theorized to be linked to periodic variations in the energy released by the impact of mass streams from the secondary star interacting with a tilted, retrogradely precessing disks \citep{1985A&A...143..313B, harvey1995superhumps, Patterson1997PASP, patterson1999permanent, 2013MNRAS.435..707A, thomas2015emergence, KimuraKIC94066522020PASJ}. Several researchers have successfully modeled NSHs based on this framework \citep{Wood2007ApJ...661.1042W, wood2009sph, Montgomery2009MNRAS, Montgomery2012}. 
In this section, we propose using NSHs as a diagnostic tool to explore the origins of the IW And-type phenomenon, integrating recent discoveries to gain further insights.

\subsection{Mass Transfer Outburst} 

Initially, variations in the mass transfer rate were proposed to explain DN outbursts through the mass-transfer burst model (MTB; \citealp{1974PASJOsaki} and \citealp{1974MNRASBath}). However, this explanation is not universally accepted.
\cite{2014mass-transferoutbursts} used simulations of V513 Cas, assuming that mass-transfer bursts followed by dips could reproduce the outbursts and dips seen in the IW And cycle (see figure 2 in \citealp{2014mass-transferoutbursts}). Despite this approach, they did not identify the underlying physical processes and suggested that the magnetic activity of the secondary star is a more likely cause. They explained the termination of standstills with outbursts (a scenario not predicted by the DIM) as a result of mass-transfer bursts from the secondary superimposed on the thermally stable state of Z Cam, with dips resulting from fluctuations in mass transfer.

In our study, we observed that the amplitude of the NSH varied with the IW And-type phenomenon in Karachurin 12. Specifically, the NSH amplitude decreased as the outburst intensified and increased as the outburst diminished. Notably, the NSH amplitude continued to rise during the ingress dip, with the maximum amplitude observed during the standstill.

It is widely accepted that the accretion disk radius expands during DN outbursts, a factor frequently considered in DN outburst models (e.g., \citealp{1992Ichikawa,1985MNRASLin,kimura2020thermal}) and crucial in explaining the superoutbursts of SU UMa systems (e.g., \citealp{1986MNRASOdonoghue,1991AcASmak,1996PASP..108...39O}). During outbursts, the disk expands due to significant angular momentum transfer to the outer edge from the accretion of large amounts of matter at the inner edge. Conversely, during quiescence, the disk's outer radius contracts as it accumulates low angular momentum mass from the secondary star’s gas stream, which has less specific angular momentum than the outer disks (e.g., \citealp{1993Ichikawa, 1984AcASmak, 2005PJABOsaki, 2020AdSpR..66.1004H}).

In the tilted disk system, the mass stream periodically impacts both sides of the disk, primarily entering the inner disks while also striking the outer edge. This phenomenon occurs because the precession timescale is more than ten times longer than the orbital period. Quantifying the impact of the mass stream on the accretion disk's radius is challenging and necessitates smoothed particle hydrodynamic (SPH) simulations (e.g., \citealp{wood2009sph}; \citealp{Montgomery2009MNRAS}; \citealp{Montgomery2012}; \citealp{thomas2015emergence}). The peak of the NSH corresponds to the innermost point that the mass stream can reach while sweeping across the disks. We assume several scenarios:

(i) The disk radius remains stable, reflecting the stability of the innermost point. In this case, mass-transfer bursts would increase the energy released by the stream, resulting in a larger NSH amplitude, which contradicts the observations in Karachurin 12.

(ii) The accretion disk expands during an outburst, leading to the outward movement of the innermost point, alongside the mass-transfer bursts. Several scenarios could influence changes in NSH amplitude here. If the disk enlarges sufficiently to surpass the effect of the mass-transfer bursts, it may lead to an decrease in NSH amplitude, and vice versa. However, understanding the interplay between changes in the accretion disks and mass transfer during this phase requires extensive simulations. We suggest that SPH can be employed to model these dynamics effectively.

(iii) Another hypothesis is that mass-transfer bursts cause the accretion disks to shrink. If a standstill is interrupted by a mass-transfer burst, as proposed by the MTB, then the energy released by the mass streams from the secondary as the outburst rises and decrease it as the outburst weakens. This scenario would predict that NSH amplitude should increase with the rising outburst and decrease as it weakens, contradicting the observations in Karachurin 12.

As noted by \cite{2014mass-transferoutbursts}, the origin of the mass transfer rate bursts remains uncertain. Additionally, the consistency of changes in NSH amplitude presents a challenge for the MTB. In summary, we contend that the MTB struggles to account for the observed phenomena in Karachurin 12.

\subsection{Tilted Thermally Unstable Disks} 

In studies of V507 Cyg, IM Eri, and FY Vul, \cite{Kato2019PASJ} proposed that the IW And-type phenomenon represents a previously unknown limit cycle oscillation. \cite{Kato2019PASJ} suggested that the standstill observed in IW And-type systems corresponds to an extended period during which the inner region of the accretion disks remains in a hot state. He also proposed that this standstill phase is eventually terminated by a thermal instability originating from the disk's outer regions, which leads to outbursts that interrupt the standstill.

Building on Kato's ideas, \cite{kimura2020thermal} explored different scenarios using a tilted, thermally unstable disk model (referred to as the tilted-DIM). Their three-dimensional hydrodynamic simulations, particularly Model B1, partially reproduced the IW And-type phenomenon under conditions of high mass transfer rates. They proposed that the tilted disk allows mass streams to enter the inner region, keeping it in a hot state for extended periods, thereby maintaining the standstill phase. Meanwhile, the outer disk remains relatively cool. Once enough matter accumulates, it triggers an outburst that interrupts the standstill. Oscillations during the standstill phase are caused by alternating cold and hot waves propagating through the middle region of the disks. A sufficiently strong cold wave reaching the inner disks can lead to a dip in brightness.

In our study of Karachurin 12, we analyzed IW And cycles indexed by dips and found diversity in their patterns:
(i) A dip is more likely to follow an outburst.
(ii) A dip can occur after a standstill.
(iii) A dip can be followed by an outburst without a preceding standstill.
(iv) Not every cycle includes a dip.

 Some of the cyclic patterns observed in Karachurin 12 are similar to the simulations of \cite{kimura2020thermal}. For example, their Model B1 shows multiple dips following a standstill, and a dip can transition directly into an outburst (see figures 11 and 12 of \citealp{kimura2020thermal}). Additionally, their Model C1 does not include a dip. We observed a significant weakening of the NSH amplitude starting from sector 73, with complete undetectability and disappearance of the IW And-type phenomenon in sectors 76 and 77. This seems to indicate a correlation between the tilted disks and the IW And-type phenomenon.

However, \cite{kimura2020thermal}'s simulations encounter similar challenges in explaining the variations in the details of IW And-type phenomenon, as they acknowledge. In Karachulin 12, dips are more likely to occur after outbursts, while in the simulations, they manifest during the quasi-standstill phase. Additionally, the amplitude of the outbursts in the simulations exceeds 1 mag. In contrast, both the outbursts and dip changes in Karachulin 12 have amplitudes of less than 1 mag. Furthermore, the simulations show dips occurring more frequently, with a maximum of six dips per cycle, a phenomenon not observed in Karachulin 12. In \cite{kimura2020thermal}'s simulations, they also estimated the precession rate, suggesting that the precession rate ($\rm v_{pre} / v_{orb}$) of the accretion disk varies with cycles. However, we found that the NSH frequency ($\rm v_{nsh} = v_{pre} + v_{orb}$) in Karachulin 12 remains relatively constant, indicating no substantial change in the precession rate.

In conclusion, while the tilted-DIM simulation partially reproduces the cyclic patterns observed in Karachulin 12, it only captures a segment of the overall morphology. Challenges persist in interpreting the detailed variations unique to Karachulin 12.

\subsection{Other Factors} 

\cite{Wood2007ApJ...661.1042W} found that an increase in the mass transfer rate also leads to a higher amplitude. \cite{Montgomery2009MNRAS} suggested that larger tilt angles of the accretion disks and large binary mass ratios are associated with greater NSH amplitudes. \cite{Wood2011ApJ} attributed variations in NSH periods to the mass distribution within the disks, indicating that an increase in inner disk mass results in longer precession and NSH periods. \cite{Osaki2013a,Osaki2013b} noted that changes in NSH periods are linked to variations in disk radius, while amplitude changes arise from additional disk components during outbursts, alongside the gas stream component. In contrast, \cite{Smak2013} proposed that larger accretion disks generally exhibit smaller NSH amplitudes, implying that fluctuations in the mass transfer rate may also play a role. \cite{thomas2015emergence} observed that the NSH amplitude increases as the viscosity parameter ($\alpha$) decreases. Consequently, describing the changes in NSHs through a single physical process proves quite challenging, as numerous factors must be considered, including mass ratio, viscosity, accretion disk tilt angle, variations in disk radius, mass distribution, and mass transfer, among others.

We discovered valuable information about the variation of NSHs with the IW And-type phenomenon, which provides valuable insights for modeling. However, we believe that using a single sample and dataset makes it difficult to comprehensively describe NSH changes; thus, large-sample, multi-data-source statistics are necessary. Additionally, time-resolved spectroscopy with adequate resolution and coordinated observations should be capable of probing the dynamics of the stream impact region, thereby enhancing our understanding of the tilted disks and IW And-type phenomenon.

\section{CONCLUSIONS} \label{sec:CONCLUSIONS}

This paper presents a detailed analysis of the newly identified IW And object, Karachurin 12, using photometric data from ASAS-SN, ZTF, and TESS. Our main findings are summarized as follows:

(1) Frequency analysis of the data reveals that the IW And cycle period for Karachurin 12 is 35.69(3) days. TESS data analysis identifies Karachurin 12 as a new NSH system with an accretion disk precession signal. We have determined the accretion disk precession period, orbital period, and NSH period for Karachurin 12 to be 4.9588(2) days, 0.3168895(13) days, and 0.2979861(8) days, respectively.
	
(2) This paper analyzes the IW And cycles in Karachurin 12 using dips as an index. A Gaussian fit to the relatively complete dip observed in the TESS photometry yields an average dip width of 3.86(10) days and a depth of 0.32(8) mag. The primary cycle patterns observed are as follows:
	(i) outburst - dip - quasi-standstill - outburst;
	(ii)outburst - quasi-standstill - outburst;
	(iii) outburst - quasi-standstill - dip - quasi-standstill - outburst;
	(iv) outburst - dip - dip - quasi-standstill - outburst;
	(v) outburst - quasi-standstill - outburst. We counted the times of different dip distances from the peak of the outburst (both pre- and post-dip), again revealing the presence of different cycles.
	These patterns highlight the diversity and complexity of IW And cycles. It is important to note that some of these cycles in Karachurin 12 have limited samples and require further validation. 
	
(3) We used the difference between the maxima and minima, segmented frequency analysis, and the Continuous Wavelet Transform method to calculate the information of NSHs. The results show that NSH amplitude decreases with outburst rise and increases with outburst recession. In the ingress dip phase NSH amplitude is allowed to continue to increase seems to continue the trend of outburst recession. No significant changes in the NSH period with the IW And cycle were observed. These phenomena indicate that the NSH serves as an ideal probe for studying IW And phenomena, offering a valuable reference for future simulations.

(4) We discuss the two dominant theories on the origin of the IW And-type phenomenon in conjunction with the phenomenon in Karachurin 12.
Mass-transfer burst model to explain the amplitude variations of NSHs observed in Karachurin 12 presents significant challenges. Typically, large mass transfers are expected to correlate with higher NSH amplitudes. Additionally, the origin of the mass-transfer bursts remains uncertain. 
The partial cycles in Karachurin 12 correspond with the findings from models of tilted thermal instability disks, but this alignment is restricted to morphological aspects. The disappearance of the IW And-type phenomenon coinciding with the undetectability of the NSHs suggests a possible connection between the tilted disks and the IW And behavior. However, there are challenges in interpreting the detailed variations in Karachurin 12, such as the timing of dips and the scales of variation and amplitude in outbursts. Notably, we observe no consistent changes in the NSH period.
Thus, we propose that a single physical process is insufficient to fully explain the variations in NSHs and the IW And-type phenomenon; instead, multiple factors must be taken into account.

\section*{Acknowledgements}

This work was supported by National Key R\&D Program of China (grant No. 2022YFE0116800), the National Natural Science Foundation of China (Nos. 11933008). We are grateful to the All-Sky Automated Survey for Supernovae for their valuable V-band photometric data of Karachurin 12, which were crucial for our analysis. Our thanks also go to the Zwicky Transient Facility for their high-cadence observations using their custom filters, which provided important temporal coverage of Karachurin 12. Additionally, we appreciate the contributions of the Transiting Exoplanet Survey Satellite for its comprehensive monitoring of variable stars, including Karachurin 12, which greatly enhanced our study. These datasets were instrumental in advancing our understanding of the IW And-type phenomenon and the behavior of Karachurin 12. All of the TESS data used in this paper can be found through the MAST: \dataset[doi:10.17909/ntst-gd72]{https://doi.org/10.17909/ntst-gd72}

\bibliography{sample631}{}
\bibliographystyle{aasjournal}

\section{Appendix}\label{sec:Appendix}
The appendix includes the following figures and tables:
 \textbf{Figure \ref{16dip}}: Displays 16 standard cycles of Karachurin 12.
\textbf{Figure \ref{fig:6dip}}: Shows three special cycles.
\textbf{Figure \ref{frequency}}: Presents the periodograms for each sector.
\textbf{Table \ref{tessLOG}}: Contains details of the TESS observations.
 \textbf{Table \ref{diptab}}: Provides information on the dips.
 \textbf{Table \ref{tab:A7}}: Lists statistics of significant signals in each sector.

\renewcommand\thefigure{A1}
\begin{figure}
	\centerline{\includegraphics[width=\columnwidth]{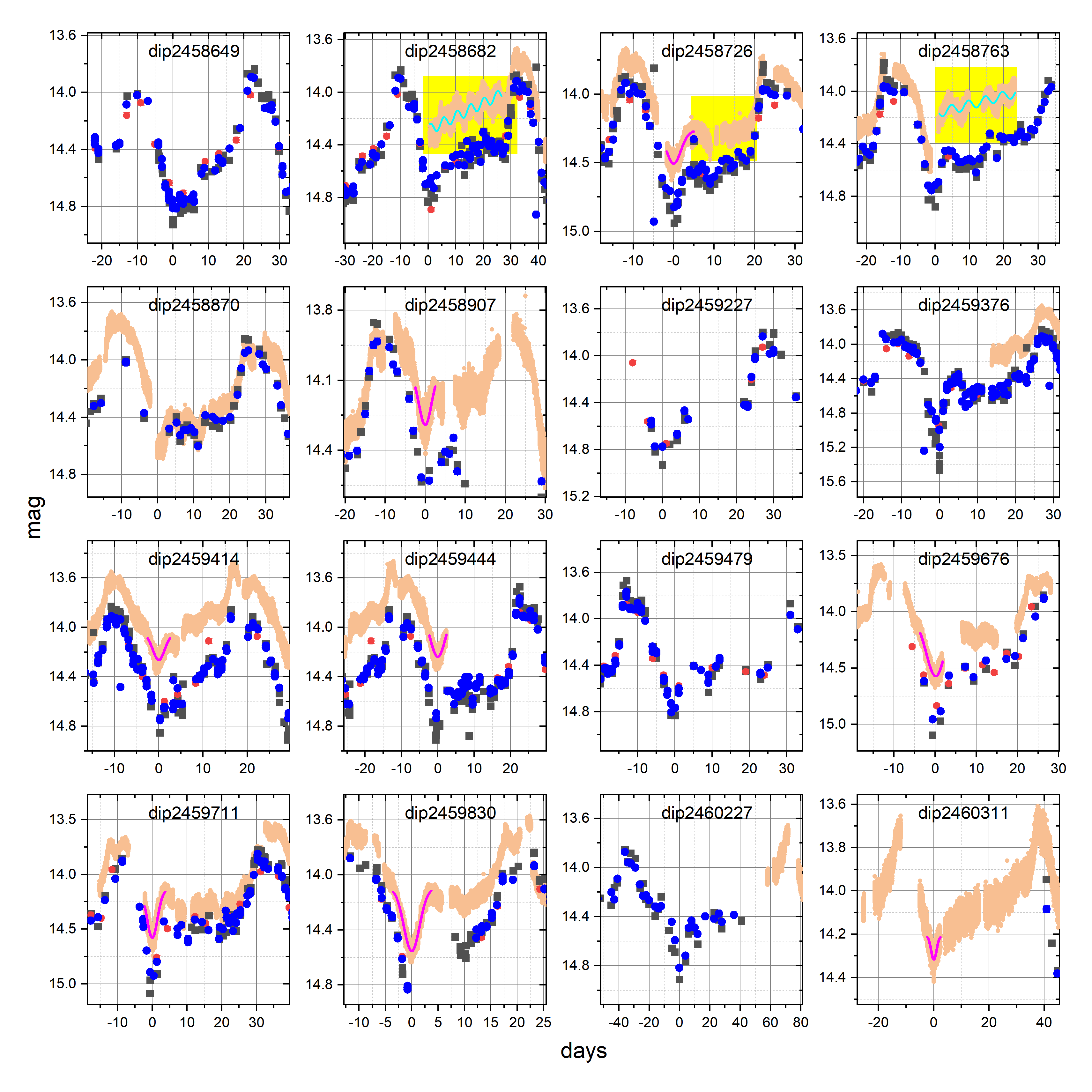}}
	\caption{Type$ \_ $I dips and their corresponding IW And cycles. For dips 2458682 and 2458763, the cyan curves on the panels represent the linear superposition of a sinusoidal fit to the precession signal. The yellow regions on the panel for dip 2458726 indicate that the precession signal is not significant. All magenta curves in the figures represent Gaussian fits to the dips.
		\label{16dip}}
\end{figure}

\renewcommand\thefigure{A2}
\begin{figure}[htbp]
\centerline{\includegraphics[width=0.6\columnwidth]{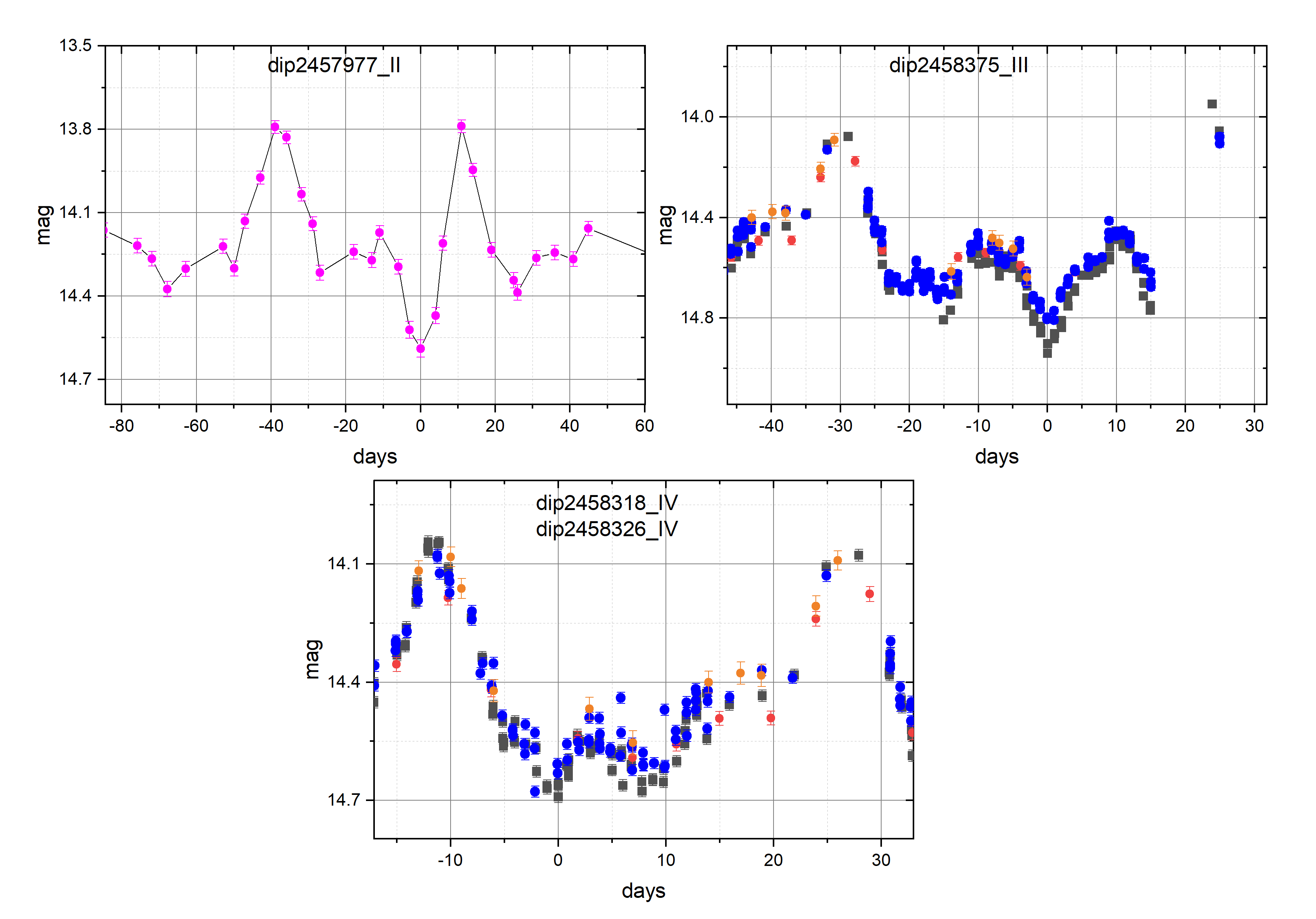}}	
	\caption{Various types of dips and their corresponding IW And cycles.}
	\label{fig:6dip}
\end{figure}

\newpage

\renewcommand\thefigure{A3}
\begin{figure*}
	\centering
	\includegraphics[width=0.45\columnwidth]{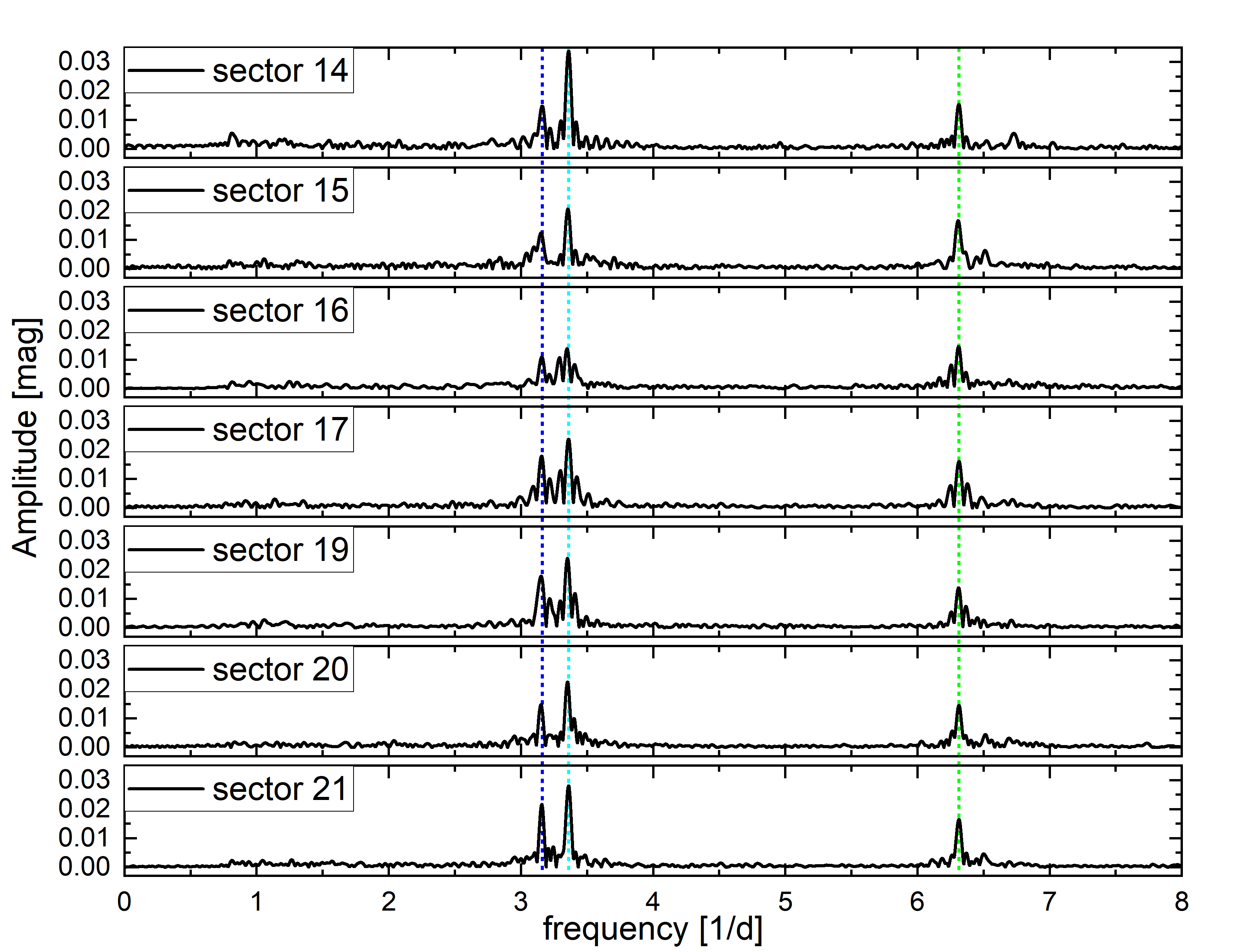}
	\includegraphics[width=0.45\columnwidth]{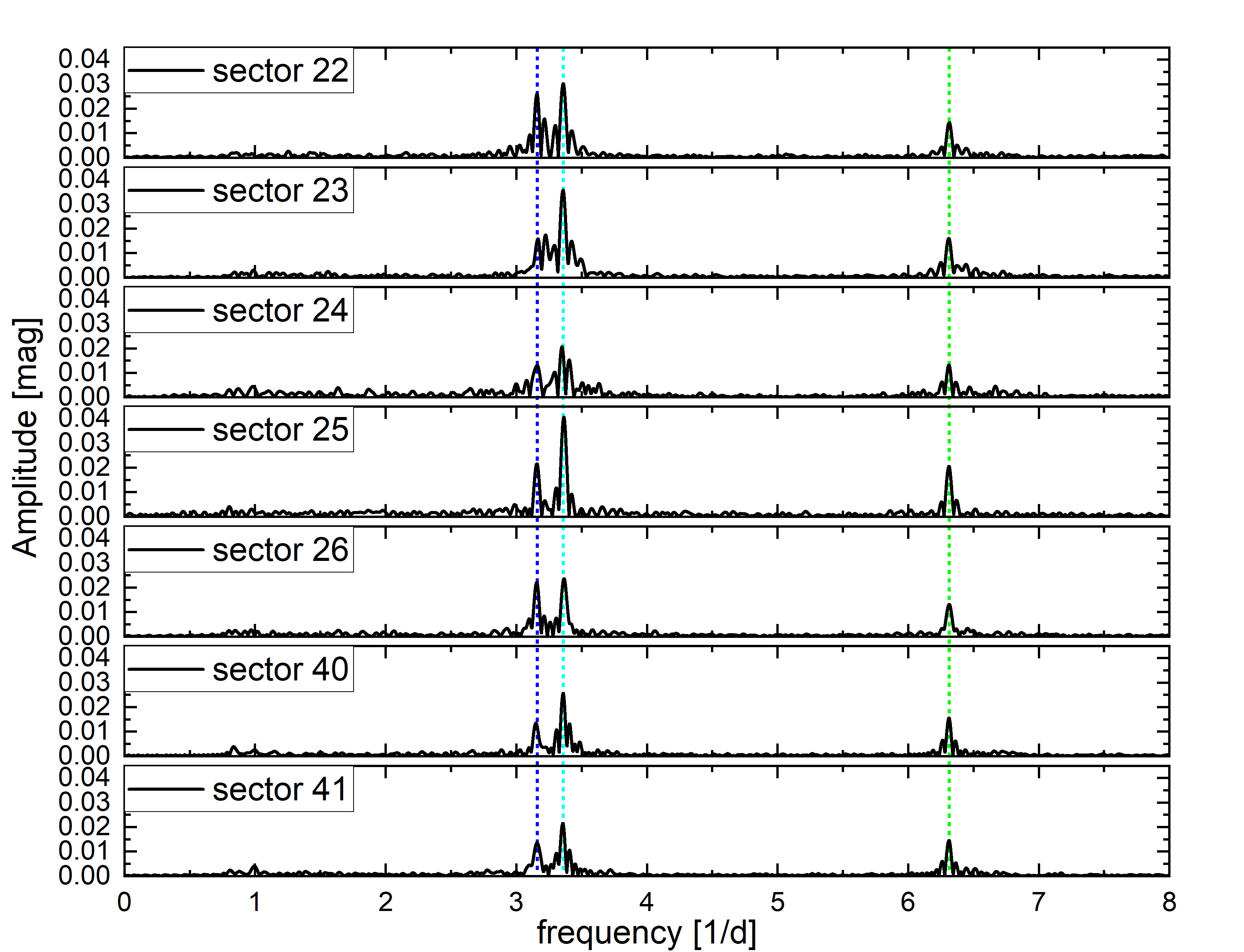}
	\includegraphics[width=0.45\columnwidth]{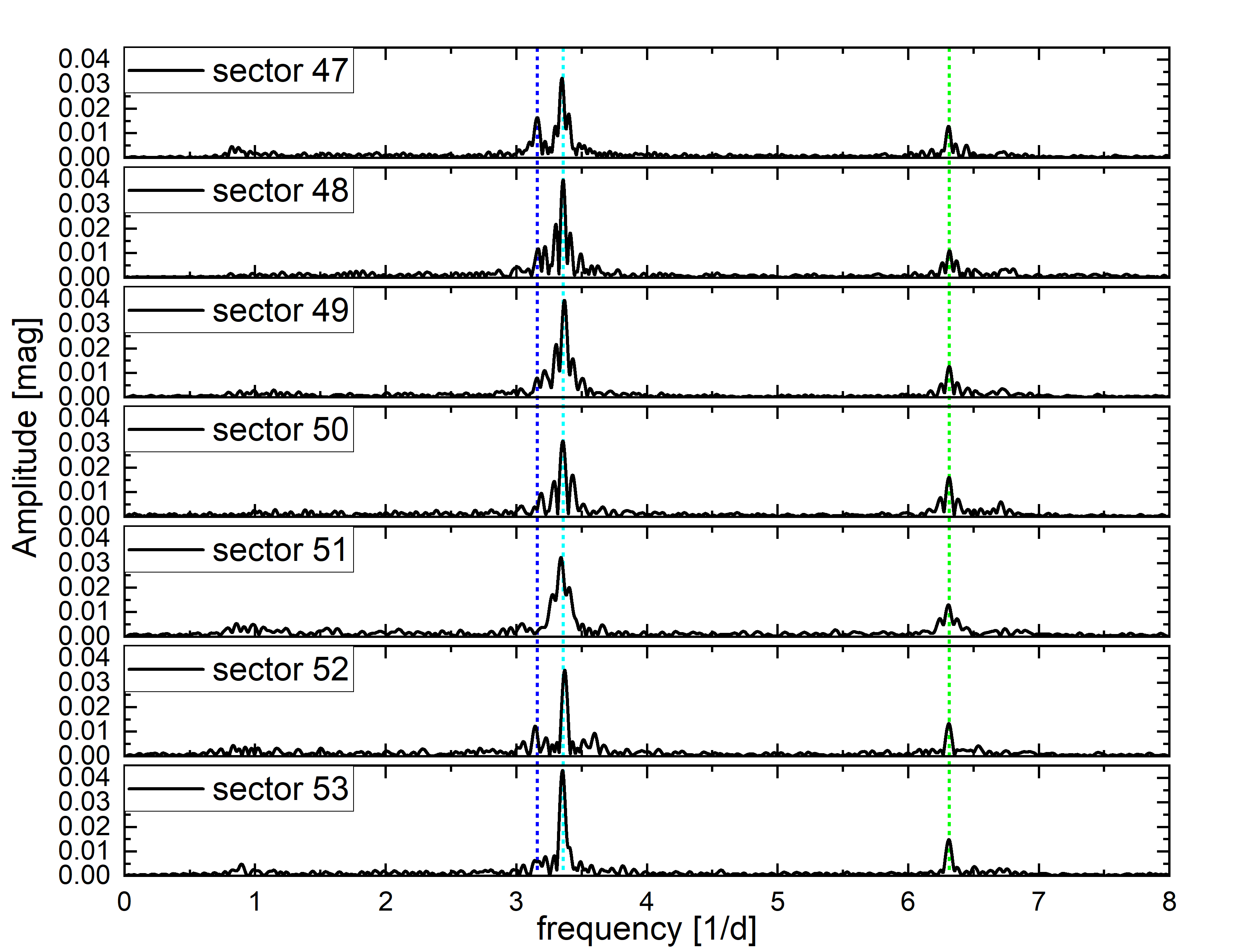}
	\includegraphics[width=0.45\columnwidth]{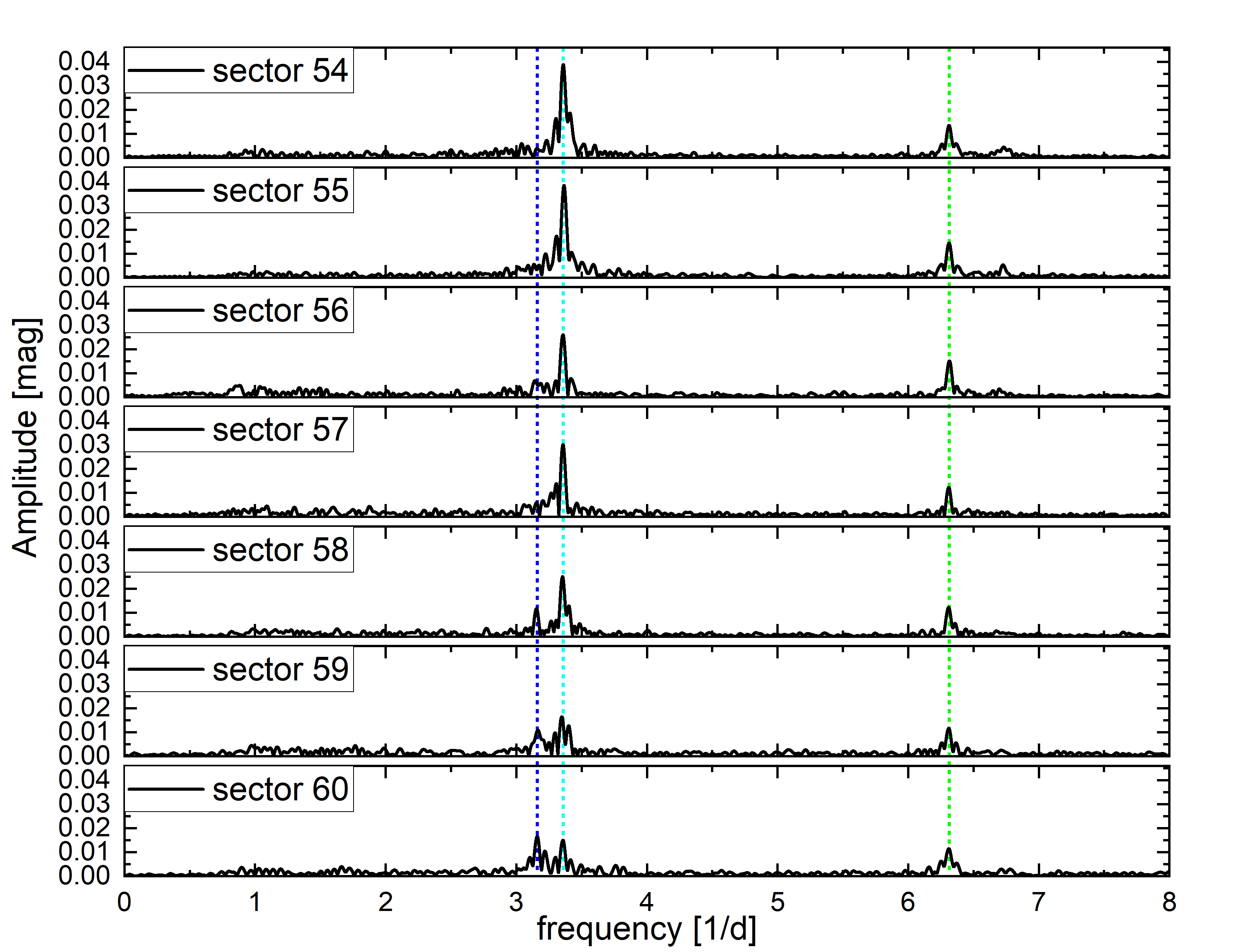}
	\includegraphics[width=0.45\columnwidth]{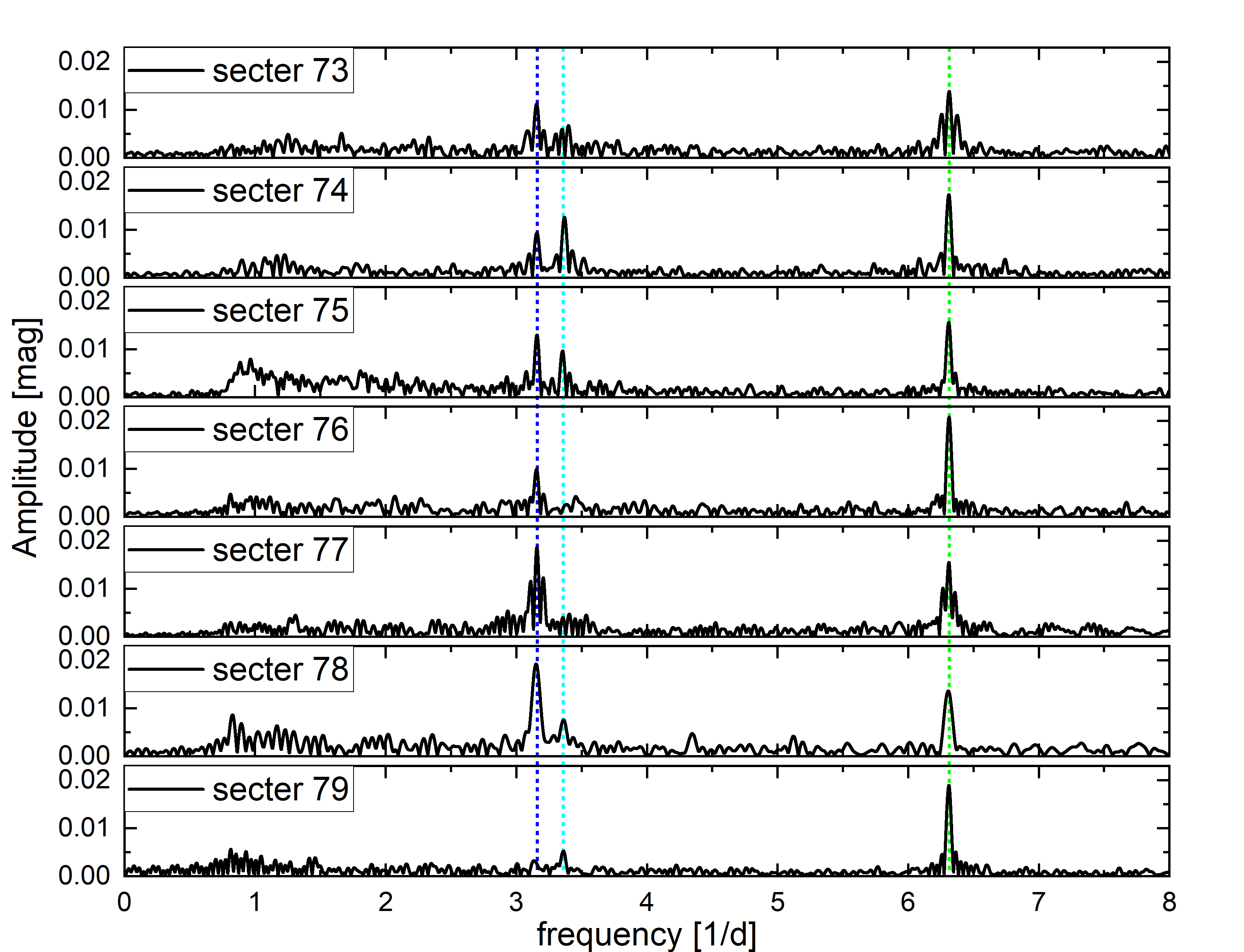}
	\caption{Periodograms corresponding to the data after removing the long-term trend in each sector. The blue, fluorescent and green vertical lines in the plot correspond to the orbital period, NSH, and double orbital period, respectively.\label{frequency}}
\end{figure*}

\newpage

\renewcommand\thetable{A1}
\begin{deluxetable*}{lll}
	\tablecaption{The TESS photometric log for Karachurin 12.  \label{tessLOG}}
	\tablewidth{1pt}
	\tabletypesize{\scriptsize}
	\tablehead{	
		\colhead{Sectors}&\colhead{Start time}&\colhead{End time}\\
	\colhead{}&\colhead{[UT]}&\colhead{[UT]}}
	\startdata
14&2019-07-18T20:27:49.737&2019-08-14T16:53:25.663\\
15&2019-08-15T20:43:24.628&2019-09-10T21:49:01.660\\
16&2019-09-12T03:37:00.636&2019-10-06T19:38:43.402\\
17&2019-10-08T04:22:43.404&2019-11-02T04:38:34.140\\
19&2019-11-28T13:58:34.620&2019-12-23T15:18:44.645\\
20&2019-12-25T00:02:45.503&2020-01-20T07:43:04.279\\
21&2020-01-21T22:23:05.569&2020-02-18T06:43:29.112 \\
22&2020-02-19T19:37:30.342&2020-03-17T23:49:51.671 \\
23&2020-03-19T14:43:52.728&2020-04-15T09:00:06.444\\
24&2020-04-16T07:04:07.261&2020-05-12T18:44:10.659 \\
25&2020-05-14T03:10:10.532&2020-06-08T19:20:03.573 \\
26&2020-06-09T18:24:01.926&2020-07-04T15:13:46.833\\
40&2021-06-25T03:39:39.725&2021-07-23T08:29:18.110\\
41&2021-07-24T11:43:16.794&2021-08-20T01:52:52.867\\
47&2021-12-31T07:30:21.863&2022-01-27T10:32:43.141\\
48&2022-01-28T10:22:43.802&2022-02-25T11:39:08.059 \\
49&2022-02-26T23:19:08.980&2022-03-25T19:33:28.773 \\
50&2022-03-26T18:29:29.248&2022-04-22T00:13:40.562\\
51&2022-04-23T10:59:40.622&2022-05-18T00:49:42.011\\
52&2022-05-19T03:13:42.212&2022-06-12T13:49:34.276\\
53&2022-06-13T11:53:32.697&2022-07-08T11:29:16.563 \\
54&2022-07-09T09:33:16.225&2022-08-04T15:04:53.810\\
55&2022-08-05T14:22:52.708&2022-09-01T18:18:28.294\\
56&2022-09-02T18:08:27.132&2022-09-30T15:18:06.137\\
57&2022-09-30T20:28:05.989&2022-10-29T14:43:53.265\\
58&2022-10-29T19:51:53.171&2022-11-26T13:07:52.610\\
59&2022-11-26T18:17:52.652&2022-12-23T04:28:02.962\\
60&2022-12-23T09:38:03.076&2023-01-18T01:58:20.973\\
73&2023-12-07T07:07:41.246&2024-01-03T03:33:55.333\\
74&2024-01-03T08:41:54.947&2024-01-30T01:38:15.948\\
75&2024-01-30T06:48:16.055&2024-02-26T23:34:39.701\\
76&2024-02-27T04:44:39.875&2024-03-25T06:38:59.495\\
77&2024-03-25T23:38:59.927&2024-04-23T01:10:32.461\\
78&2024-05-03T18:30:33.351&2024-05-21T19:44:32.759\\
79&2024-05-22T00:54:32.687&2024-06-18T04:10:21.917\\
	\enddata
	\begin{threeparttable}
	\end{threeparttable}
\end{deluxetable*}

\newpage
\renewcommand\thetable{A2}
\begin{deluxetable*}{lllllllllllll}
	\tablecaption{Parameters for different dips.\label{diptab}}
	\tablewidth{0.01pt}
	\tabletypesize{\scriptsize}
	\tablehead{    
		\colhead{Name} & \colhead{Type} &\colhead{HJD} & \colhead{mag} & \colhead{err} & \colhead{FWHM} & \colhead{err} & \colhead{Depth} &\colhead{err} & \colhead{distance$ \_ $1$ ^{\rm a} $} & \colhead{distance$ \_ $2$ ^{\rm b} $} & \colhead{source}
	}
	\startdata
dip2457977&II&2457977.806 &14.589 &0.031 &-&-&-&-&38.97 &10.96 &ASAS$\_$SN\\
dip2458284&III&2458284.872 &14.798 &0.014 &-&-&-&-&37.05 &22.98 &ZTF$\_$zg\\
dip2458318&IV&2458318.907 &14.691 &0.014 &-&-&-&-&11.06 &27.92 &ZTF$\_$zg\\
dip2458326&IV&2458326.700 &14.677 &0.014 &-&-&-&-&-&-&ZTF$\_$zg\\
dip2458375&III&2458375.719 &14.940 &0.014 &-&-&-&-&28.89 &-&ZTF$\_$zg\\
dip2458649&I&2458649.865 &14.929 &0.014 &-&-&-&-&9.94 &22.96 &ZTF$\_$zg\\
dip2458682&I&2458682.681 &14.833 &0.014 &-&-&-&-&9.86 &32.04 &ZTF$\_$zg\\
dip2458726&I&2458726.661 &14.504 &0.033 &4.547 &0.068 &0.259 &0.036 &11.94 &22.10 &TESS\\
dip2458763&I&2458763.681 &14.880 &0.014 &-&-&-&-&14.91 &32.96 &ZTF$\_$zg\\
dip2458870&I&2458870.816 &14.684 &0.028 &-&-&-&-&12.72 &24.24 &TESS\\
dip2458907&I&2458907.974 &14.292 &0.040 &3.699 &0.163 &0.228 &0.052 &12.92 &22.18 &TESS\\
dip2459057&II&2459057.759 &16.177 &0.016 &-&-&-&-&28.91 &13.99 &ZTF$\_$zg\\
dip2459227&I&2459227.033 &14.936 &0.014 &-&-&-&-&-&27.00 &ZTF$\_$zg\\
dip2459336&IV&2459336.923 &14.847 &0.014 &-&-&-&-&12.03 &29.03 &ZTF$\_$zg\\
dip2459344&IV&2459344.953 &14.812 &0.014 &-&-&-&-&-&-&ZTF$\_$zg\\
dip2459376&I&2459376.893 &15.460 &0.015 &-&-&-&-&10.94 &26.89 &ZTF$\_$zg\\
dip2459414&I&2459414.504 &14.264 &0.035 &3.176 &0.083 &0.213 &0.041 &10.68 &17.44 &TESS\\
dip2459444&I&2459444.189 &14.239 &0.035 &4.754 &0.299 &0.356 &0.069 &12.24 &22.54 &TESS\\
dip2459479&I&2459479.658 &14.835 &0.014 &-&-&-&-&12.93 &30.97 &ZTF$\_$zg\\
dip2459676&I&2459676.681 &15.098 &0.014 &5.301 &0.092 &0.515 &0.049 &13.22 &27.07 &TESS\\
dip2459711&I&2459711.545 &14.578 &0.051 &3.558 &0.038 &0.443 &0.055 &7.79 &32.71 &TESS\\
dip2459830&I&2459830.459 &14.554 &0.035 &3.516 &0.021 &0.460 &0.037 &11.64 &23.20 &TESS\\
dip2460227&I&2460227.716 &14.913 &0.014 &-&-&-&-&-&-&ZTF$\_$zg\\
dip2460311&I&2460311.310 &14.316 &0.283 &2.346 &0.059 &0.110 &0.285 &11.98 &37.81 &TESS\\
	\enddata
	\begin{threeparttable}
	\begin{tablenotes}[para,flushleft]
		\item [a] The time from the previous outburst peak to the dip.\\
		\item [b] The time from the dip to the next outburst.
	\end{tablenotes}
\end{threeparttable}	
	
\end{deluxetable*}

\newpage
\clearpage

\renewcommand\thetable{A3}

\setlength{\tabcolsep}{5pt}
\begin{longtable}{ccccccccccccc}

	\caption{Frequency analysis results for different sectors.}\label{tab:A7}\\

	%\hline
	
	\toprule
	Sector&f$\rm  _{orb} $&err&amp.&err&f$\rm _{nsh} $&err&amp.&err&f$\rm _{5} $&err&amp.&err\\

&[1/d]&[1/d]&[mag]&[mag]&[1/d]&[1/d]&[mag] &[mag]&[1/d]&[1/d]&[mag]&[mag]\\
	\midrule
	\endfirsthead
	\multicolumn{13}{c}
	{{\bfseries{Table \thetable\ }continued from previous page}} \\
	\toprule
	Sector&f$\rm  _{orb} $&err&amp.&err&f$\rm _{nsh} $&err&amp.&err&f$\rm _{5} $&err&amp.&err\\
&[1/d]&[1/d]&[mag]&[mag]&[1/d]&[1/d]&[mag] &[mag]&[1/d]&[1/d]&[mag]&[mag]\\
\midrule
	\endhead
	\midrule
	\multicolumn{13}{c}
	{{ }} \\
	\endfoot
	\endlastfoot

ALL&3.156 &1.3E-05&0.009 &4.1E-04&3.356 &8.9E-06&0.014 &4.1E-04&6.311 &8.6E-06&0.015 &4.1E-04\\
14&3.159 &1.0E-02&0.018 &3.4E-04&3.360 &6.1E-03&0.034 &3.8E-04&6.313 &1.2E-02&0.016 &3.3E-04\\
15&3.154 &1.4E-02&0.011 &3.0E-04&3.355 &9.0E-03&0.021 &3.4E-04&6.307 &1.0E-02&0.017 &3.1E-04\\
16&3.156 &1.4E-02&0.011 &2.8E-04&3.348 &1.1E-02&0.014 &2.9E-04&6.311 &1.1E-02&0.015 &3.0E-04\\
18&3.155 &1.0E-02&0.017 &3.1E-04&3.360 &7.8E-03&0.024 &3.4E-04&6.313 &9.9E-03&0.016 &2.9E-04\\
19&3.154 &8.3E-03&0.018 &2.8E-04&3.352 &6.9E-03&0.024 &3.0E-04&6.311 &1.0E-02&0.014 &2.6E-04\\
20&3.152 &1.0E-02&0.014 &2.5E-04&3.352 &7.0E-03&0.023 &2.8E-04&6.313 &1.0E-02&0.015 &2.6E-04\\
21&3.157 &7.7E-03&0.020 &2.7E-04&3.360 &5.8E-03&0.028 &3.0E-04&6.314 &8.5E-03&0.016 &2.5E-04\\
22&3.157 &6.0E-03&0.028 &3.0E-04&3.359 &5.9E-03&0.030 &3.2E-04&6.314 &1.0E-02&0.014 &2.7E-04\\
23&3.161 &1.0E-02&0.016 &3.0E-04&3.357 &5.3E-03&0.036 &3.5E-04&6.310 &1.1E-02&0.016 &3.2E-04\\
24&3.152 &1.4E-02&0.015 &3.7E-04&3.349 &1.0E-02&0.021 &3.8E-04&6.311 &1.5E-02&0.013 &3.5E-04\\
25&3.157 &9.9E-03&0.019 &3.4E-04&3.364 &5.7E-03&0.041 &4.2E-04&6.311 &9.7E-03&0.021 &3.7E-04\\
26&3.156 &7.7E-03&0.022 &3.0E-04&3.365 &7.5E-03&0.024 &3.2E-04&6.313 &1.2E-02&0.013 &2.9E-04\\
40&3.153 &8.7E-03&0.015 &2.3E-04&3.358 &5.7E-03&0.026 &2.7E-04&6.312 &8.6E-03&0.016 &2.4E-04\\
41&3.158 &9.3E-03&0.014 &2.4E-04&3.355 &6.8E-03&0.022 &2.7E-04&6.312 &9.3E-03&0.015 &2.5E-04\\
47&3.159 &1.1E-02&0.016 &3.0E-04&3.349 &5.7E-03&0.033 &3.3E-04&6.309 &1.2E-02&0.013 &2.8E-04\\
48&3.161 &1.7E-02&0.011 &3.3E-04&3.358 &5.2E-03&0.040 &3.8E-04&6.315 &1.6E-02&0.011 &3.2E-04\\
49&3.150 &1.7E-02&0.009 &2.7E-04&3.368 &4.7E-03&0.040 &3.3E-04&6.314 &1.3E-02&0.013 &3.0E-04\\
50&3.177 &2.5E-02&0.006 &2.8E-04&3.356 &6.0E-03&0.031 &3.4E-04&6.312 &1.0E-02&0.016 &3.0E-04\\
51&-&-&-&-&3.341 &7.4E-03&0.032 &4.3E-04&6.307 &1.6E-02&0.013 &3.7E-04\\
52&-&-&-&-&3.370 &6.5E-03&0.035 &4.1E-04&6.312 &1.5E-02&0.013 &3.7E-04\\
53&-&-&-&-&3.353 &4.5E-03&0.043 &3.5E-04&6.311 &1.1E-02&0.015 &3.1E-04\\
54&-&-&-&-&3.353 &4.5E-03&0.043 &3.5E-04&6.311 &1.1E-02&0.015 &3.1E-04\\
55&-&-&-&-&3.359 &5.2E-03&0.039 &3.7E-04&6.312 &1.3E-02&0.013 &3.1E-04\\
56&3.145 &3.1E-02&0.005 &2.9E-04&3.357 &6.9E-03&0.026 &3.3E-04&6.315 &1.1E-02&0.016 &3.1E-04\\
57&3.148 &3.1E-02&0.007 &4.2E-04&3.357 &8.4E-03&0.030 &4.6E-04&6.310 &1.9E-02&0.012 &4.3E-04\\
58&3.156 &2.3E-02&0.009 &3.8E-04&3.354 &9.3E-03&0.025 &4.2E-04&6.309 &1.8E-02&0.012 &4.0E-04\\
59&3.164 &2.6E-02&0.011 &5.1E-04&3.347 &1.8E-02&0.017 &5.4E-04&6.310 &2.5E-02&0.012 &5.3E-04\\
60&3.160 &2.0E-02&0.017 &6.0E-04&3.356 &2.4E-02&0.014 &5.8E-04&6.310 &2.7E-02&0.012 &5.7E-04\\
73&3.155 &2.4E-02&0.011 &4.9E-04&3.349 &4.9E-02&0.006 &4.8E-04&6.314 &2.0E-02&0.014 &5.1E-04\\
74&3.159 &2.5E-02&0.010 &4.4E-04&3.368 &2.0E-02&0.013 &4.5E-04&6.311 &1.5E-02&0.017 &4.7E-04\\
75&3.158 &2.0E-02&0.013 &4.8E-04&3.354 &2.9E-02&0.009 &4.7E-04&6.310 &1.7E-02&0.016 &4.9E-04\\
76&3.154 &2.8E-02&0.010 &5.0E-04&-&-&-&-&6.313 &1.4E-02&0.021 &5.2E-04\\
77&3.158 &1.6E-02&0.019 &5.5E-04&-&-&-&-&6.311 &1.9E-02&0.015 &5.3E-04\\
78&3.151 &1.7E-02&0.019 &5.8E-04&3.357 &4.0E-02&0.008 &5.5E-04&6.306 &2.3E-02&0.014 &5.7E-04\\
79&-&-&-&-&3.360 &3.7E-02&0.005 &3.6E-04&6.311 &1.1E-02&0.019 &3.8E-04\\

	\bottomrule
\end{longtable}	
\footnotesize{Note:} The notations f$\rm _{orb}$, f$\rm _{nsh}$, and f$\rm _{5}$ refer to orbital frequencies, NSH frequencies, and second orbital harmonics, respectively.

\end{document}